\documentclass[draftcls,12pt,onecolumn]{IEEEtran}

\usepackage[a-2b]{pdfx} 

\usepackage{amsmath,amssymb,amsfonts}
\usepackage{bm}
\usepackage{mathtools}
\usepackage{nicefrac}
\usepackage{array}
\usepackage{blkarray}
\usepackage[symbol]{footmisc}

\usepackage{graphicx}
\usepackage{epsfig}
\usepackage{float}
\usepackage{multirow}
\usepackage{subcaption}
\usepackage{blkarray}
\usepackage[font=small,skip=0pt]{caption}

\usepackage{xcolor}
\usepackage{soul}
\usepackage{changes}

\usepackage{cite}
\usepackage{hyperref}

\begin{document}

\title{\large \textbf{LDGM-Based Quantum Codes for Fault-Tolerant Quantum Computation}\thanks{This material is based upon work supported by the National Science Foundation under Grant CCF-2007689. The work was developed while all authors where at the University of Delaware.}}

\author{
    Yumin Li$^1$, Kejing Liu$^2$, Hanqing Lou$^3$, and Javier Garcia-Frias$^1$ \\
    \vspace{0.1em}
    \small{
    $^1$Department of Electrical and Computer Engineering, University of Delaware;
    $^2$Verizon;
    $^3$InterDigital\\
    Email:{\{liyumin,jgf\}@udel.edu}}
}

\date{}

\maketitle
\renewcommand{\thefootnote}{\arabic{footnote}}
\thispagestyle{empty}

\vspace{-1cm}

\begin{abstract}
We construct a new family of Calderbank-Shor-Steane (CSS) codes using 
the generator and parity-check matrices of Low-Density Generator Matrix (LDGM) codes, with row operations applied 
to both matrices in order to achieve the desired quantum rate. Decoding is 
performed in an iterative manner, by applying message passing over 
the associated graph, and discrete Density Evolution (DDE) is used to optimize performance in the depolarizing channel. The proposed construction offers high flexibility and easiness in the design, producing quantum codes that possess excellent error correction capabilities. By properly designing the structure of the code, we are able to control and bound the weight of the stabilizer generators to a small value, which results in codes particularly well suited for fault-tolerant quantum computation. At the same time, these codes achieve very good performance in terms of error correction capability.

\end{abstract}

\normalsize

\section{Introduction}\label{sec:intro}

Quantum error-correcting codes and fault-tolerant quantum computation have drawn much interest during the
last years. The first proposed schemes dealt with very short block
lengths, and the error correction capabilities of these codes were
far away from the achievable limits \cite{shor95,NCbook}. An
important class of quantum error-correcting codes, known as CSS
codes, was proposed by Calderbank, Shor and Steane in
\cite{cs96,steane96}. The CSS framework comprises a large number of
particular quantum error-correcting codes and provides a unifying
perspective. Even a more generalized class, the class of stabilizer
codes, was proposed in \cite{gottesmanThesis,CRSS,crss98}, where it was shown that
quantum error-correcting codes can be constructed based on classical
block codes.

While these developments in quantum error correction were taking
place, classical error-correcting codes were also developing at a
very fast pace. The discovery of capacity approaching codes with
reasonable encoder/decoder computational complexity, such as turbo
codes \cite{berrou} and low-density parity check (LDPC) codes
\cite{gallager,mackay1999}, was one of the most important
developments in coding during the 1990's. Turbo-like codes have
been successfully applied in many different communications problems, 
and also to construct quantum error correcting codes.

The first quantum codes based on LDPC codes appeared in
\cite{postol,mackay,camara,spawc05}. Of special interest were the
results in \cite{mackay}, where CSS sparse-graph codes
making use of iterative decoding were introduced. The codes in
\cite{mackay} were  based on classical LDPC codes with specific
constraints required by the CSS construction, and
easily outperformed existing quantum codes, since large block
lengths could be constructed. Decoding was performed by using iterative
techniques. The quantum rate could be easily adjusted, but the
code structure utilized in \cite{mackay} to satisfy the CSS constraints led to some performance degradation.

Motivated by \cite{mackay}, in \cite{spawc05,turbo06,ciss08} we
proposed novel quantum error-correcting codes based on the use of
the generator and parity-check matrices of LDGM codes. Classical
LDGM codes \cite{gz2003,asil,Wei} are a special class of LDPC codes
with a sparse generator matrix. Therefore, they can be
decoded with the same complexity as standard LDPC codes, but the
encoding complexity is much less. The rationale of the proposed
quantum error-correcting codes is that in systematic LDGM codes the
generator matrix and the parity check matrix are both low density
and orthogonal to each other. This provides a high degree of
flexibility in the construction of quantum codes, since the CSS
constraints can be easily satisfied by using these matrices. Importantly for fault-tolerant quantum computation, the stabilizer generators of these codes have small (bounded) weight and so do the encoded $\bar{X}$ and $\bar{Z}$ operators. This is very relevant, since the number of physical gates required to realize encoded Clifford operators is highly dependent on the weight of $\bar{X}$ and $\bar{Z}$ \cite{Rengaswamy,Rengaswamy2}. In particular, the number of physical CNOTs required to implement logical gates increases with the weight, which amplifies error propagation and raises the probability of gate failures during quantum computation.

Recent work in quantum LDPC codes has studied constructions that are asymptotically optimal and achieve a minimum distance that increases linearly with the block length \cite{Panteleev_SIGACT,Panteleev_IT,golowich}.
Thus, the choice of LDGM codes may at first appear questionable, as their minimum distance is, by definition, small
and independent of the block length, which will lead to error floors.
However, as explained by MacKay in \cite{mackay}, “Much of coding theory, including the founding papers of quantum coding theory, emphasizes the number of errors that can be corrected by a code--an emphasis which leads to one trying to make codes that have large minimum distance. However, distance is not everything”, and the minimum distance does not need to increase with the block length for a code to correct (asymptotically) almost all of the errors up to the theoretical limit: quoting \cite{mackay} again “if our goal is simply to make a code whose error probability is smaller than some figure such as $10^{-6}$ or $10^{-20}$ then there is no reason why the distance has to grow at all with blocklength''. Indeed, as in classical LDGM codes \cite{gz2003,asil,Wei,liu_isit10,chai05,isit05}, we will see in the sequel that the trade-off between convergence threshold and error floor appearing in our proposed quantum LDGM codes can be kept under control to obtain excellent performance in terms of error rate, even when they are designed for fault-tolerant computation.

The objective of this paper is to construct simple random-like LDGM-based quantum codes with excellent error correction capabilities, and which can be successfully used for fault-tolerant quantum computation. In order to achieve this goal, we expand the work in \cite{spawc05,turbo06,ciss08} in several directions: i) we show how exploiting the correlation between the binary representation of the Pauli errors introduced by a depolarizing channel leads to substantial improvements in performance, ii) we introduce an optimization technique based on the use of
Discrete Density Evolution (DDE) \cite{dde}, which facilitates the design of better codes, and iii) more importantly, we design codes that not only show excellent error correction capabilities but are also perfectly suited for fault-tolerant computation.

The remainder of this paper is as follows. Section II provides a brief overview of stabilizer and CSS codes, and presents the key idea that allows the use of classical LDGM codes to generate CSS codes, discussing also how decoding is performed. Section III describes the proposed structure in detail and Section IV proposes a novel design to utilize LDGM quantum codes for fault-tolerant computation. Section V studies the decoding modifications required to deal with the depolarizing channel, while Discrete Density Evolution (DDE) is developed in Section VI. Simulation results illustrating the excellent performance of the proposed codes are provided in Section VII. Finally, Section VIII concludes the paper.

\section{Quantum Error-Correcting Codes Based on LDGM Codes}
\label{sec:quantumldgm}

\subsection{Classical Low-Density Generator Matrix Codes}\label{sec:ldgm}

Systematic LDGM codes are linear codes with a sparse generator matrix
$\tilde{G}=[I\ P]$, so that the output bits $o=[o_1 \dots o_L]^T$ are obtained from the input bits, $u=[u_1 \dots u_K]^T$ as $o^T=u^T\tilde{G}=[u^T \; \; u^TP]$.  
Since LDGM codes are a specific class of LDPC codes---their parity-check matrix $\tilde{H}=[P^T\ I]$ is also sparse---they can be
decoded in exactly the same manner, using message passing algorithms such as belief propagation \cite{pearl1988} or the sum-product algorithm \cite{kschischang2001}. These algorithms operate on graphs representing the parity-check matrix or the equation $c=P^Tu$, where ${c}$ is the
vector of parity bits generated at the encoder, $P$ is the non-systematic part
of the generator matrix, and ${u}$ is the information message. This framework also accounts for non-systematic LDGM codes, where the input bits are not part of the output bits, i.e., $\tilde{G}=P$. During decoding, message passing utilizes prior knowledge of $u$, noisy observations of $c$ ($c'$), and---for systematic codes---noisy observations of $u$ ($u'$). Depending on the ratio between $K$ and $L$ and the presence of input bits in the output, an LDGM code can function as either a channel code or a source code.

We define $(x,y)$ regular LDGM codes as those where the degrees of the input and parity bit nodes are $x$ and $y$, respectively. 
It is well known that systematic regular LDGM codes are asymptotically bad
\cite{mackay1999}. This was corroborated in \cite{gz2003,asil},
where it was shown that the use of single regular LDGM codes always
leads to error floors which do not improve with block length. However, the trade-off between convergence threshold and error floor can be managed by properly selecting the node degrees. Moreover, a significant reduction in the error floor
can be achieved by serially concatenating two
regular LDGM codes \cite{gz2003,asil} or by using a parallel concatenation scheme
\cite{chai05,isit05}. The resulting performance for concatenated
LDGM codes over classical BSC and AWGN channels is, as shown in
\cite{gz2003,asil,chai05}, comparable to that of irregular LDPC and
turbo codes, but with a very low encoding/decoding complexity. 

\subsection{Stabilizer Codes}

Stabilizer codes \cite{gottesmanThesis} are defined by a set of $N-K$ ``stabilizer generators"
$S=\{ S_i \}$, which is a set of independent\footnote{Independent in the sense that none of them can be expressed as a product of other generators (up to a global phase).} Pauli operators on $N$ qubits such that any two operators in the set commute and  $-I^{\otimes N} \notin S$. A quantum codeword is defined as a state $| \psi \rangle$ that
is a $+1$ eigenstate of all the stabilizer generators (i.e., $S_i | \psi \rangle = | \psi \rangle$ for all $i$.) The set of all codewords, or codespace, is a subspace of dimension $2^K$. Due to the properties of the Pauli
operators, a given Pauli error operator $E_{\alpha}$ on $N$ qubits either commutes or
anticommutes with each stabilizer generator $S_i$. Then, the
syndrome of $E_{\alpha}$ is defined as the ``commutation status''
(either commute or anticommute) of $E_{\alpha}$ with respect to all
the generators, and is completely independent of the quantum state
$| \psi \rangle$.

A Pauli operator on $N$ qubits can be written uniquely as a product
of an $X$-containing operator (i.e., using only Pauli matrices $X$ and
$I$), a $Z$-containing operator, and a phase factor $(+1,-1,i$ or $-i)$. Then, discarding the phase, the $X$- ($Z$-) containing operator can be expressed as a
binary string of length $N$, with `1' standing for X (Z) and `0' for
I. For instance, $XIYZYI=-(XIXIXI)\times (IIZZZI)=(101010|001110)$.
In this way, we can represent each generator stabilizer as a binary
vector, and write the set of generators as a binary matrix
$A=(A_1|A_2)$, where row $j$ of $A_1$ ($A_2$) corresponds to the
$X$- ($Z$-) containing operator of stabilizer generator $j$. With
this binary representation, the commutativity between stabilizers
can be expressed as

\begin{equation}
A_1A_2^T + A_2A_1^T = 0 \pmod 2. \label{twisted}
\end{equation}

Similarly, a Pauli error operator $E'_{\alpha}$ can be
interpreted as a binary string $e'_{\alpha}=\left(%
\begin{array}{c}
  e^x \\
  e^z \\
\end{array}%
\right)$
of length $2N$.  By
exchanging the $X$ and $Z$ operators in $E'_{\alpha}$, the
ordinary dot product (mod 2) of $e_{\alpha}=\left(%
\begin{array}{c}
  e^z \\
  e^x \\
\end{array}%
\right)$ with a row of the
matrix $A$ is 0 if $E'_{\alpha}$ and the stabilizer represented by
that row commute, and 1 otherwise. Thus, the quantum syndrome for
the error operator $E' _{\alpha}$ is exactly the classical syndrome
$Ae_{\alpha}$, where matrix $A=(A_1|A_2)$, called quantum parity
check matrix, acts as the classical parity check matrix and
$e_{\alpha}$ as the binary error pattern. Therefore, we can conclude
that from any binary matrix $(A_1|A_2)$ of size $M_Q \times 2N$
satisfying (\ref{twisted}), it is possible to construct a quantum code that encodes $K=N-M_Q$ qubits in $N$ physical qubits.

\subsection{CSS Codes}

Calderbank-Shor-Steane (CSS) codes \cite{cs96,steane96} are an important subfamily of stabilizer codes. Their quantum parity-check matrix has the form

\begin{equation}
A=(A_1|A_2)=\left(%
\begin{tabular}{c|c}
  H & 0 \\
  0 & G \\
\end{tabular}
\right),
\label{css}
\end{equation}
with $GH^T=0$, which implies $HG^T=0$. Notice that this assumption guarantees that condition
(\ref{twisted}) is always satisfied in CSS codes, and thus CSS codes are a special class of
stabilizer codes. CSS constructions
based on the use of appropriate classical codes can be found in
\cite{almeida,forney,grassl1,grassl2,grassl3,hagiwara,ollivier}.

\subsection{Construction of LDGM-Based CSS Codes}\label{sec:ldgmCSS}

From the previous development, it is clear that in order to build
``good" quantum codes we just need to design ``good" classical
binary codes with parity-check matrix $A=(A_1|A_2)$ satisfying (\ref{twisted}). Notice that this is different from the design
of classical LDPC-like codes, since (\ref{twisted}) imposes
important constraints in the code design. The first quantum codes
based on LDPC codes appeared in \cite{postol,mackay,camara,spawc05}.
Of special interest are the results in \cite{mackay}, where classical LDPC
codes were used to construct matrix $A$, and the proposed codes
easily outperformed previous quantum codes. Specifically, the
quantum parity check matrix of these codes, named Dual-Containing
Low-Density Parity-Check codes (DC-LDPC), satisfies $G=H$ in
(\ref{css}), with the constraints that i) every row in $H$ has
weight $k$ and every column has weight $j$; ii) every pair of rows
in $H$ has an even overlap, and every row has even weight. These
constraints guarantee that (\ref{twisted}) is satisfied.

The quantum rate of DC-LDPC codes can be easily adjusted, but the
restrictions introduced by their special (dual-containing) matrix
structure lead to significant performance degradation with respect
to classical LDPC codes with optimized parity-check matrices. The idea we proposed in \cite{spawc05,turbo06}
is to allow more degrees of freedom for the quantum parity check
matrix $A$ by using the generator and the parity-check matrix of
classical systematic LDGM codes in CSS codes. As explained before,
the rationale for utilizing LDGM codes is that both the parity check
matrix $\tilde{H}$ and the generator matrix $\tilde{G}$ of an LDGM
code are sparse and easy to build. More importantly, $\tilde{H}$ and
$\tilde{G} $ are orthogonal by definition (i.e., $\tilde{G}
\tilde{H}^T =0$), which forces the
compliance of ($\ref{twisted}$).

The first intuitive idea would be to directly use $\tilde{H}$ and
$\tilde{G}$ in (\ref{css}). However, the resulting matrix
$\tilde{A}$ would have size $N \times 2N$, and, therefore, it could
not be used for encoding purposes, since the resulting quantum rate
would be $0$. In order to get a valid quantum code, we need to reduce the
number of rows, while keeping the constraint (\ref{twisted}). The simplest way to do this is to puncture several rows of
matrix $\tilde{G}$. However, the systematic structure of $\tilde{G}$
makes it impossible to decode successfully in this case, since
puncturing $\tilde{G}=[I\ P]$ leads to a new matrix where some
columns have degree $0$. The technique we proposed in
\cite{spawc05,turbo06} to construct quantum codes from matrix
$\tilde{A}$ is to perform linear row operations on both $\tilde{G}$
and $\tilde{H}$ to reduce the number of rows. The following theorem details the process.

\textbf{Theorem} \emph{Let $\tilde{H}=[P^T\ I]$ of size $n_1 \times
N$ and $\tilde{G}=[I\ P]$ of size $n_2 \times N$ be the parity-check
matrix and the generator matrix for a given classical linear block
code C, respectively (i.e., $n_1+n_2=N$). Define $H=M_1 \tilde{H}$
and $G=M_2 \tilde{G}$, where $M_1$ has size $m_1 \times n_1$ such
that $m_1<n_1$, and the size of $M_2$ is $m_2 \times n_2$ such that
$m_2<n_2$. The matrix
$$ A=\left(%
\begin{array}{c|c}
  H & 0 \\
  0 & G \\
\end{array}%
\right) $$ of size $(m_1+m_2) \times 2N$ is the parity-check matrix of a quantum CSS code with quantum
rate $R_Q=\frac{K}{N}$, where $K=N-m_1-m_2$. }\\
\textbf{Proof:}
Since $\tilde{G}\tilde{H}^T= 0$, 
$GH^T
=
(M_2 \tilde{G})(M_1 \tilde{H})^T
=
M_2 (\tilde{G}\tilde{H}^T) M_1^T
=
0.$
Therefore, the CSS constraint is preserved, and the quantum rate can be adjusted by appropriately selecting matrices $M_1$ and $M_2$.




\subsection{Decoding of LDGM-Based CSS
Codes}\label{sec:decodeldgmCSS}

\begin{figure}[!t]
\centering \resizebox{3.0in}{!}{
\includegraphics{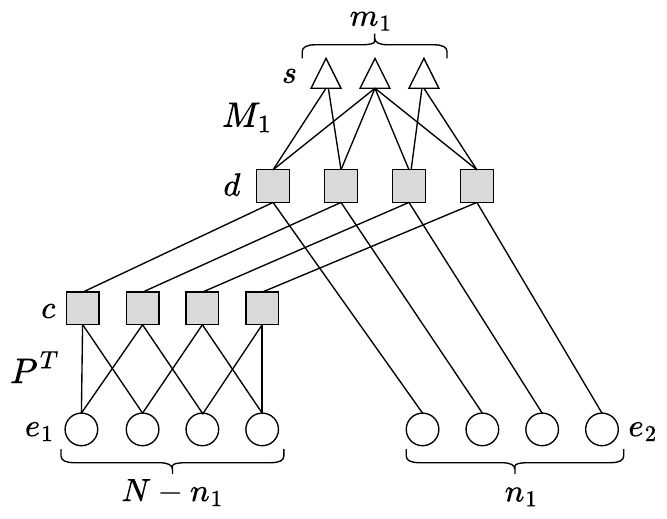}}
\caption{Decoding graph for matrix $H$, consisting of the $X$-containing stabilizer generators, utilized to decode $e^z$ ($Z$-decoder) if errors $e^x$ and $e^z$ are independent. To simplify notation, we denote the syndrome $s^x$ by $s$ and the error pattern $e^z$ by $e$.} 
\label{fig:decoding}
\end{figure}

Decoding is performed by applying the belief propagation/sum-product
algorithm \cite{pearl1988,kschischang2001} on the graph defined by
matrix $A$
\begin{equation}
\left(%
\begin{array}{c}
  s^x \\
  s^z \\
\end{array}%
\right)=
\left(%
\begin{array}{c|c}
  H & 0 \\
  0 & G \\
\end{array}%
\right)
\left(%
\begin{array}{c}
  e^z \\
  e^x \\
\end{array}%
\right),
\end{equation}
where $s^x$ is the vector of syndromes measured using the $X$-containing
stabilizer generators defined in matrix $H$ and $s^z$ is the vector of syndromes measured using the $Z$-containing stabilizer generators defined in matrix $G$.
Notice that if errors $e^x$ and $e^z$ are independent, decoding for the $H$ and $G$
matrices can be performed independently on the graphs defined by $s^x=He^z$ ($Z$-decoder) and  $s^z=Ge^x$ ($X$-decoder), respectively. Fig. \ref{fig:decoding}
represents the decoding graph for matrix $H$, consisting of the $X$-containing stabilizer generators, utilized to decode $e^z$, i.e., the $Z$-decoder. A similar process is performed for matrix $G$, consisting of the $Z$-containing stabilizer generators, to decode $e^x$ ($X$-decoder). Focusing on matrix $H$, to simplify notation in Fig. \ref{fig:decoding} we denote the syndrome by $s$ and the error pattern by $e$. They are related by $s=He=M_1 \tilde{H} e=M_1 [P^T\
I]e$. In order to construct the graph in Fig. \ref{fig:decoding}, we
split the error pattern $e$ into two parts $e_1$ and $e_2$, and relate
it to the syndrome in a two-step process

\begin{equation}
d=[P^T \ I]e=[P^T \ I]\left(%
\begin{array}{c}
  e_1 \\
  e_2 \\
\end{array}%
\right)=P^T e_1+e_2
\label{d}
\end{equation}

\begin{equation}
s=M_1d.
\label{s}
\end{equation}

The section in the lower part of the graph (lower layer) in Fig. \ref{fig:decoding}, where information is exchanged between nodes $e_1$ and $c$, and between nodes $e_2/c$ and $d$, is defined by (\ref{d}) and can be seen as a classical non-systematic LDGM code performing source coding to generate parity bits $d$ from input $e$ when the non-systematic part
of the generator matrix (as defined in Section II.A) is $[P^T \ I]^T$. On the other hand, the upper part of the graph (upper layer) in Fig. \ref{fig:decoding}, specified in detail the sequel (Section III.A, Fig. \ref{figure1}), is defined by (\ref{s}) and can be seen as a classical LDPC code with syndromes $s$ and parity bits $d$.

For decoding, we apply belief propagation over the graph defined in Fig.
\ref{fig:decoding}. Notice that the syndrome $s$ is known exactly
from the quantum decoding circuit, and the {\it a priori}
probability of the error pattern $e$ can be calculated according to
the quantum channel. However, there is no information about the
middle layer of the graph. In other words, we do not have any {\it a
priori} information about variables $c$ or $d$, which introduces
difficulties in the first decoding iterations. In order to
overcome this problem, we utilize the \emph{doping} \cite{doping}
technique in matrix $M_1$. That is, we introduce some
degree-1 syndrome nodes, which propagate correct information to some
nodes of $d$, and push the iterative decoding in the right
direction. It is important to mention that decoding based on the
graph presented in Fig. \ref{fig:decoding} is different from
decoding based on the final matrix $H$. The reason is that the
product of $M_1$ and $\tilde{H}$ eliminates some edges and
introduces more cycles\footnote{Notice that the decoding structure based on
Fig. \ref{fig:decoding} is similar to a serial concatenated LDGM
scheme in classical error correction, where decoding of the overall
code also results in worse performance \cite{Wei}.}, thus degrading the performance of iterative decoding methods.

\section{Proposed Structure}\label{sec:structure}

In order to simplify the code design, we force matrix $P^T$ to have the same degree distribution for nodes $c$ and $e_1$ (obviously, this implies $n_1=n_2=N/2$, see Fig.
\ref{fig:decoding}), so that the degree distribution of $P$ is the
same as that of $P^T$. In this way, the results for the decoding of
$e^z$ errors using matrix $H$ ($Z$-decoder) are statistically the same as the results
obtained for the decoding of $e^x$ errors using matrix $G$ ($X$-decoder),
and, if errors $e^x$ and $e^z$ are independent (something that does not occur for a depolarizing channel), we just need to simulate either the $X$-decoder or the $Z$-decoder. For matrices $M_1$ and $M_2$, we fix $M_1=M_2=M$ so that $m=m_1=m_2=\frac{N-K}{2}$. As mentioned above and described in detail next, we apply doping.


\subsection{Proposed Upper Layer Structure: Doping}\label{sec:lowdegree nondoping ldgmCSS}
\begin{figure}[!t]
\centerline{\includegraphics[width=0.6\textwidth,height=0.2\textheight]{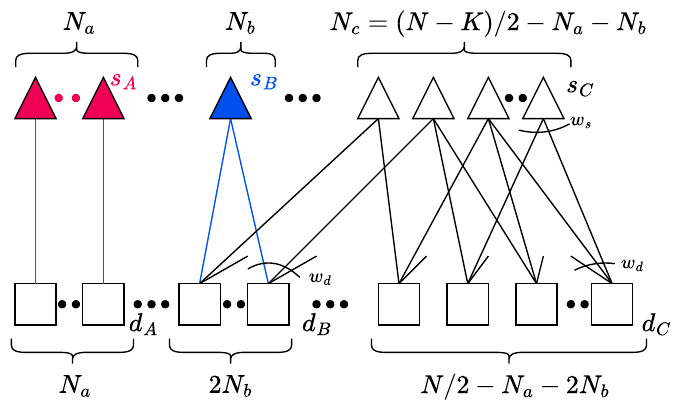}}
\caption{Graph associated with proposed matrix $M$. Each doping/$s_A$ syndrome node is connected to one $d$ node ($d_A$ node). $s_B$ syndrome nodes have degree two, while all $s_C$ syndrome nodes have degree $w_s$. Each $d_B$ node is connected to only one $s_B$ node, while $d_C$ nodes are connected only to nodes $s_C$. All $d_B$ and $d_C$ nodes have degree $w_d$. For a fixed value of $K$ and $N$, the parameters to optimize are $w_d$ and the number of $s_A$ and $s_B$ nodes ($N_a$ and $N_b$, respectively). $w_s$ and the number of $s_C$, $d_A$, $d_B$ and $d_C$ nodes are fixed given the parameters above.}
\label{figure1}
\end{figure}

Fig. \ref{figure1} shows the proposed upper layer structure (matrix $M$). As mentioned above, we utilize \emph{doping} \cite{doping}, introducing $N_a$ degree-1 syndrome nodes in the associated graph, referred to as doping syndrome nodes or $s_A$ syndrome nodes. These nodes propagate ``correct" messages to the $d$ nodes they are connected to ($d_A$ nodes), pushing the iterative decoding in the right direction during the first decoding iterations. Notice that for a given $d_A$ node connected to a specific $s_A$ node, the messages propagated downwards to the $c$ and $e_2$ nodes are independent of the messages that node $d_A$ receives from other non-doping syndrome nodes. Moreover, messages passed from a $d_A$ node to non-doping syndrome nodes will never lead to an improved performance for the iterative decoder. Therefore, as shown in Fig. \ref{figure1} a $d_A$ node should only be connected to an $s_A$ node.

Notice that there is an important design trade-off in the number of doping/$s_A$ syndrome nodes: If there are too few $s_A$ nodes, the iterative decoding process lacks sufficient ``correct" information to be pushed in the right direction during the first decoding iterations, and in that sense it is desirable for the number of doping syndrome nodes to be high. However, if too many $s_A$ nodes are introduced, the degree of the remaining non-doping syndrome nodes has to increase significantly if the density of 1's in matrix $M$ (number of edges in the graph associated with matrix $M$) is fixed. This makes the messages coming from these high-degree non-doping syndrome nodes less reliable, which negatively affects the performance of the iterative decoder.
The choice of the number of doping/$s_A$ syndrome nodes is, therefore, a
trade-off between these two effects.

To address this trade-off, as shown in Fig. \ref{figure1}, our proposed structure utilizes $N_b$ non-doping syndrome nodes of degree 2, $s_B$ nodes. Notice that the degree of $s_B$ nodes is higher than that of $s_A$ nodes, but lower than that of the $N_c$ remaining syndrome nodes ($s_C$ nodes, all with degree $w_s$). Therefore, $s_B$ nodes do not convey messages as correctly as degree-1 $s_A$ doping nodes, but their messages are more accurate than those of $s_C$ nodes.  Utilizing a combination of $s_A$, $s_B$, and $s_C$ syndrome nodes provides more flexibility in code design and softens the aforementioned trade-off. For instance, if $N_a+N_b$ is fixed, and the overall number of connections between $s$ and $d$ nodes is also fixed, increasing the value of $N_b$ results in a smaller value for $w_s$, increasing the accuracy of the messages proceeding from the $s_C$ nodes.

To ensure that the more reliable messages proceeding from $s_B$ nodes propagate to the highest possible number of $d$ nodes, we do not allow any node $d$ to be connected to more than one $s_B$ node. Thus, each one of the $2N_b$ $d_B$ nodes connected to $s_B$ nodes will receive one relatively reliable message from a $s_B$ node among its $w_d$ incoming messages. As we will see in the sequel, this structure of matrix $M$ will have to be modified in order to design codes well suited for fault-tolerant quantum computation.

\subsection{Proposed Lower Layer Structure: Regular LDGM
Codes}\label{sec:structure_regular}

As explained before, $\tilde{G}$ and $\tilde{H}$ are, respectively, the
generator and parity-check matrices of a regular LDGM code. For
simplicity, we just consider regular matrices $P$ in Fig.
\ref{fig:decoding} with size $\frac{N}{2} \times \frac{N}{2}$  ($n_1=\frac{N}{2}$) and degrees $(y,y)$.
An important characteristic of the proposed structure is the presence of error floors that do not
decrease with the block length. These error floors were analytically studied for very simple quantum channels in \cite{ciss09}.

\section{Design of Codes Well-Suited for Fault-Tolerant Quantum Computation}
In order to develop good fault-tolerant schemes, it is important to be able to implement logical gates with few physical gates. In particular, it is desirable to design codes so that the $K$ encoded/logical $\bar{Z}$ operators and the $K$ encoded/logical $\bar{X}$ operators have few positions different from $I$. Equivalently, the binary representation of these operators, $\bar{Z}$ and $\bar{X}$, both of dimension $K \times 2N$, should have low weight for all their rows. 

It is well known that the parity-check matrix of any $[N,K]$ CSS code can be expressed, through Gaussian elimination and simultaneous column permutation in the X and Z sections, as

\begin{equation}
\makeatletter
\setlength{\BA@colsep}{3pt}
\makeatother
\renewcommand{\arraystretch}{1}
\begin{blockarray}{c@{\hspace{10pt}}cccccc}
  & \overbrace{\hphantom{I}}^{\frac{N-K}{2}}
  & \overbrace{\hphantom{A_1}}^{\frac{N-K}{2}}
  & \overbrace{\hphantom{A_2}}^{K}
  & \overbrace{\hphantom{D}}^{\frac{N-K}{2}}
  & \overbrace{\hphantom{I}}^{\frac{N-K}{2}}
  & \overbrace{\hphantom{E}}^{K}
  \\[-5pt]
\begin{block}{c[ccc|ccc]}
\frac{N-K}{2}\{ & I & A_1 & A_2 & 0 & 0 & 0 \\
\frac{N-K}{2}\{ & 0 & 0 & 0 & D & I & E \\
\end{block}
\end{blockarray} \; ,
\label{standardformCSS}
\end{equation}
so that the encoded/logical $\bar{X}$ operators and the encoded/logical $\bar{Z}$ operators are obtained as

\begin{equation}
    \begin{aligned}
        \bar{X} &= 
        \left[
        \begin{array}{ccc|ccc}
        0 & E^T & I & 0 & 0 & 0
        \end{array}
        \right] \\[4pt]
        \bar{Z} &= 
        \left[
        \begin{array}{ccc|ccc}
        0 & 0 & 0 & A_2^T & 0 & I
        \end{array}
        \right].
    \end{aligned}
    \label{encodedXZ}
\end{equation}

For the CSS codes proposed in the previous section with $M_1=M_2=M$, the parity-check matrix is given as

\begin{equation}
        A = 
        \left[
        \begin{array}{cc}
        M & 0 \\
        0 & M 
        \end{array}
        \right]
        \left[
        \begin{array}{cc|cc}
        P^T & I & 0 & 0\\
        0 & 0 & I & P 
        \end{array}
        \right] \\[4pt]
        = 
        \left[
        \begin{array}{cc|cc}
        M P^T & M& 0 & 0\\
        0 & 0 & M & M P 
        \end{array}
        \right],
    \label{intermediateH}
\end{equation}
where the matrix $P$ can be defined as

\begin{equation*}
\begin{aligned}
P \;=\;
&\begin{blockarray}{c@{\hspace{5pt}}cc}
  & \overbrace{\hphantom{P_{11}}}^{\frac{N-K}{2}}
  & \overbrace{\hphantom{P_{12}}}^{\frac{K}{2}}
  \\[-5pt]
\begin{block}{c[cc]}
\makebox[2.5em][r]{$\frac{N-K}{2}\lbrace$} & P_{11} & P_{12} \\
\makebox[2.5em][r]{$\frac{K}{2}\lbrace$}   & P_{21} & P_{22} \\
\end{block}
\end{blockarray}.
\\[3pt]
\end{aligned}
\end{equation*}

\begin{figure}[]
\centerline{\includegraphics[width=0.65\textwidth,height=0.18\textheight]{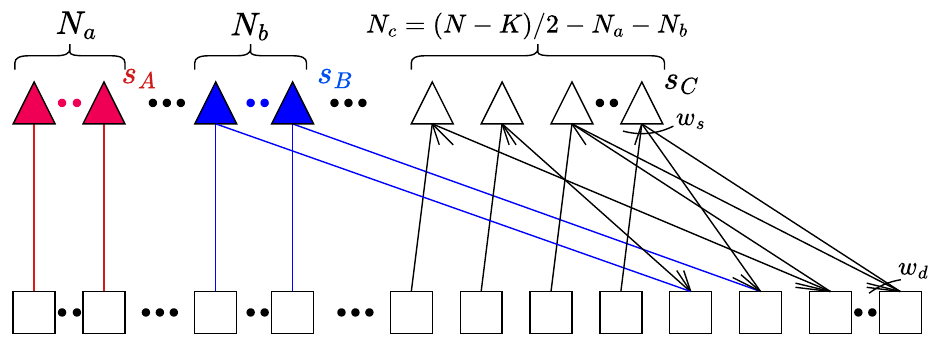}}
\caption{Graph associated with the matrix $M$ utilized in the upper layer for the design of codes well suited for fault-tolerant quantum computation. In the lower layer, matrix $P$ is modified by forcing $P_{22}=0$.}
\label{figure3}
\end{figure}

Different from the previous section, for reasons that will be clear shortly we modify the structure of matrix $M$, which is now defined as

\begin{equation*}
    M = \left[
    \begin{array}{cc}
    I & M'
    \end{array}
    \right], \; \; \;
    M' =
    \begin{array}{r}
    N_a\{ \\
    \frac{N-K}{2}-N_a\{
    \end{array}
    \left[
    \begin{array}{c}
    0\\
    M''
    \end{array}
    \right],
\end{equation*}
so that
\begin{equation}
    M = \left[
    \begin{array}{ccc}
    I & 0 & 0\\
        0 & I & M'' 
    \end{array}
    \right],
    \label{matrixM}
\end{equation}
where the column weight of matrix $M'$ is $w_d$, and $M'$ has additional constraints as defined in its corresponding graph shown in Fig. \ref{figure3}.

Substituting (\ref{matrixM}) into (\ref{intermediateH}), the parity-check matrix of the proposed fault-tolerant CSS codes takes the following form

\begin{equation*}
\makeatletter
\setlength{\BA@colsep}{3pt} 
\makeatother
\renewcommand{\arraystretch}{1}
\begin{blockarray}{c@{\hspace{6pt}}cccccccc}
  & \overbrace{\hphantom{P_{11}^T+M'P_{12}^T}}^{\frac{N-K}{2}}
  & \overbrace{\hphantom{P_{21}^T+M'P_{22}^T}}^{\frac{K}{2}}
  & \overbrace{\hphantom{I}}^{\frac{N-K}{2}}
  & \overbrace{\hphantom{M'}}^{\frac{K}{2}}
  & \overbrace{\hphantom{I}}^{\frac{N-K}{2}}
  & \overbrace{\hphantom{M'}}^{\frac{K}{2}}
  & \overbrace{\hphantom{P_{11}+M'P_{21}}}^{\frac{N-K}{2}}
  & \overbrace{\hphantom{P_{12}+M'P_{22}}}^{\frac{K}{2}}
  \\[-5pt]
\begin{block}{c[cccc|cccc]}
\frac{N-K}{2}\{ & P_{11}^T+M'P_{12}^T
      & P_{21}^T+M'P_{22}^T
      & I
      & M'
      & 0 & 0 & 0 & 0 \\
\frac{N-K}{2}\{ & 0 & 0 & 0 & 0
      & I
      & M'
      & P_{11}+M'P_{21}
      & P_{12}+M'P_{22} \\
\end{block}
\end{blockarray} \; .
\label{standardformproposed_alt}
\end{equation*}



We now aim to transform this parity-check matrix into the standard form given in (\ref{standardformCSS}). To achieve this, it suffices to permute certain columns of the matrix. However, to preserve the symplectic structure of the CSS code, any column permutation applied to the $X$ part must also be applied to the corresponding columns of the $Z$ part. Therefore, we move the third column blocks of both parts to the first column position to obtain the standard form of the proposed code as

\begin{equation}
\makeatletter
\setlength{\BA@colsep}{3pt}
\makeatother
\renewcommand{\arraystretch}{1}
\begin{blockarray}{c@{\hspace{6pt}}cccccccc}
  & \overbrace{\hphantom{I}}^{\frac{N-K}{2}}
  & \overbrace{\hphantom{P_{11}^T+M'P_{12}^T}}^{\frac{N-K}{2}}
  & \overbrace{\hphantom{P_{21}^T+M'P_{22}^T}}^{\frac{K}{2}}
  & \overbrace{\hphantom{M'}}^{\frac{K}{2}}
  & \overbrace{\hphantom{P_{11}+M'P_{21}}}^{\frac{N-K}{2}}
  & \overbrace{\hphantom{I}}^{\frac{N-K}{2}}
  & \overbrace{\hphantom{M'}}^{\frac{K}{2}}
  & \overbrace{\hphantom{P_{12}+M'P_{22}}}^{\frac{K}{2}}
  \\[-5pt]
\begin{block}{c[cccc|cccc]}
{(N-K)/2} \{ & I
      & P_{11}^T+M'P_{12}^T
      & P_{21}^T+M'P_{22}^T
      & M'
      & 0 & 0 & 0 & 0 \\
{(N-K)/2} \{ & 0 & 0 & 0 & 0
      & P_{11}+M'P_{21}
      & I
      & M'
      & P_{12}+M'P_{22}\\
\end{block}
\end{blockarray} \; .
\label{standardformproposed}
\end{equation}


By comparing (\ref{standardformproposed}) with (\ref{standardformCSS}), we identify the corresponding submatrices in (\ref{standardformCSS}) as 
\begin{align*}
A_1 &=  P_{11}^T + M' P_{12}^T \\
A_2 &= \left[ P_{21}^T + M' P_{22}^T \;\; M' \right] \\
D   &= P_{11} + M' P_{21} \\
E   &= \left[ M' \;\; P_{12} + M' P_{22} \right] .
\end{align*}

The parity-check matrix of the proposed fault-tolerant CSS codes is obtained by substituting these submatrices into (\ref{standardformCSS}), and making $P_{22}=0$, which introduces additional constraints in the general structure discussed in the previous section. From (\ref{encodedXZ}), the encoded/logical $\bar{X}$ and $\bar{Z}$ operators are then

\begin{equation*}
    \begin{aligned}
        \bar{X} &=
        \left[
        \begin{array}{cccc|cccc}
        0 & M'^T & I & 0 & 0 & 0 & 0 & 0\\
        0 & P_{12}^T & 0 & I & 0 & 0 & 0 & 0
        \end{array}
        \right] \\[4pt]
        \bar{Z} &=
        \left[
        \begin{array}{cccc|cccc}
        0 & 0 & 0 & 0 & P_{21} & 0 & I & 0 \\
        0 & 0 & 0 & 0 & M'^T & 0 & 0 &  I
        \end{array}
        \right].
    \end{aligned}
\end{equation*}

Notice that the weight of half of the encoded/logical $\bar{X}$ operators is determined by the row weight of submatrix $P_{12}^T$ or, equivalently, by the column weight of submatrix $P_{12}$ which is $y$ when we consider a regular LDGM code of degrees $(y,y)$ (recall that $P_{22}=0$). Thus, the weight of this half of the $\bar{X}$ operators will always be $y+1$. 
The weight of the other half of the encoded/logical $\bar{X}$ operators is determined by the row weight of matrix $M'^T$ or, equivalently, by the column weight of matrix $M'$, which as explained before is $w_d$, the degree of the nodes $d$ that are connected to more than one syndrome node (see Fig. \ref{figure3}). Thus, the weight of this other half of the $\bar{X}$ operators will be $w_d+1$. Notice that the reason $M$ was defined as specified in Fig. \ref{figure3} was to fix the weight of half of the $\bar{X}$ operators to $w_d+1$, where $w_d$ can be seen as a design parameter that has to be chosen properly. For instance, if $w_d=y$, the weight of all the encoded/logical $\bar{X}$ operators is $y+1$. A similar argument holds for the encoded/logical $\bar{Z}$ operators.


\section{Modified decoder for the depolarizing channel}
\begin{figure}[!t]
\centering \resizebox{3in}{!}{
\includegraphics{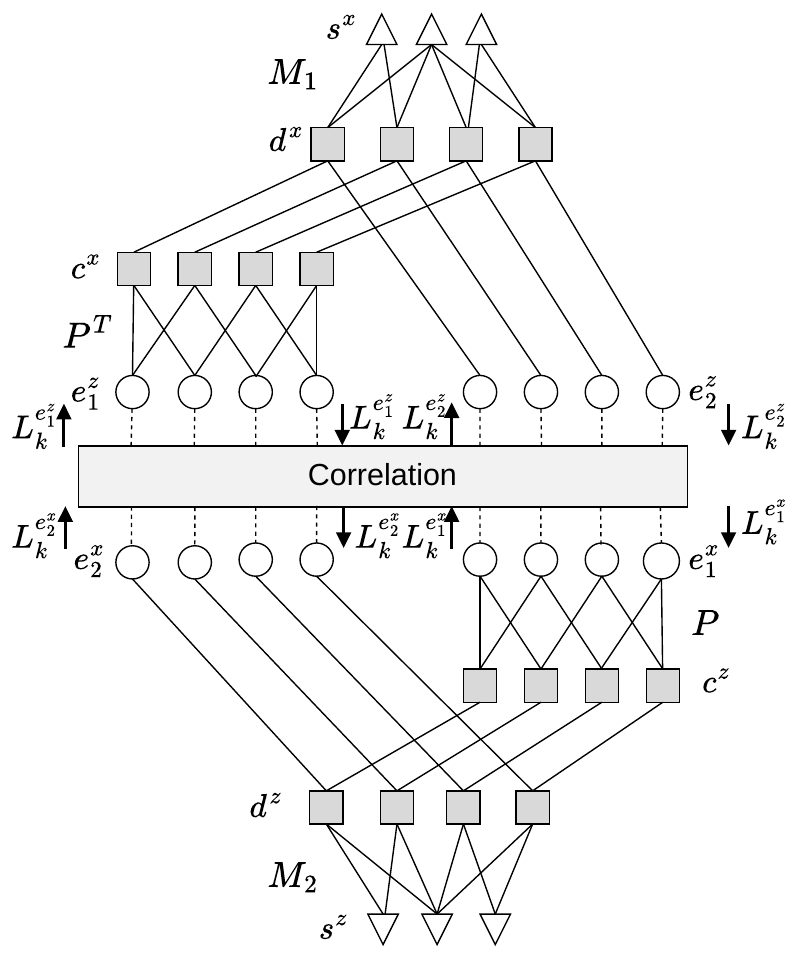}}
\caption{Decoding graph for a CSS code with matrices $H=M_1 [P^T \; I]$ and $G=M_2 [I \; P]$ utilized in the depolarizing channel. The upper part of the graph ($Z$-decoder) and the lower part of the graph ($X$-decoder) exchange information to exploit the correlation existing between errors $e^z$ and $e^x$. In our development, $M_1=M_2=M$.}
\label{fig:decodergraphdep}
\end{figure}

Our focus in this paper is on the depolarizing channel, where Pauli errors $X$, $Y$ and $Z$ each occur with probability $p/3$, and $p$ is the depolarizing parameter. Thus, the errors $e^x$ and $e^z$ are correlated. Therefore, in order to optimize
performance\footnote{If the correlation is ignored, decoding can be performed independently in the $X$- and the $Z$-decoders as defined in Fig. \ref{fig:decoding}. The price to pay will be performance degradation, which we will assess in the sequel.} this correlation has to be exploited in the decoding process.
The key is to iteratively exchange information between the $X$-decoder and the
$Z$-decoder based on the graph depicted in Fig. \ref{fig:decodergraphdep}, in a manner very similar to the case of distributed turbo-like coding
of correlated sources in classical coding, an area pioneered by one of the authors \cite{gz2003,dcc01,tcom_noise,cl,spm07}.

Specifically, the message passed from the $k^{th}$ node of the $X$-decoder, $e_{i,k}^x$, to
the $k^{th}$ node of the $Z$-decoder, $e_{\hat{i},k}^z$,
where $i,\hat{i} \in \{1,2\}, \hat{i}=3-i$, is given by
\begin{equation}\label{eqn:p_corr}
\begin{aligned}
p_k^{e_{\hat{i}}^z}
&\propto p_k^{e_i^x} P(e^z=1,e^x=1)
 + (1-p_k^{e_i^x})P(e^z=1,e^x=0) \\
&= \frac{p}{3}p_k^{e_i^x} + \frac{p}{3}(1-p_k^{e_i^x})
\\[6pt]
1-p_k^{e_{\hat{i}}^z}
&\propto p_k^{e_i^x} P(e^z=0,e^x=1)
 + (1-p_k^{e_i^x})P(e^z=0,e^x=0) \\
&= \frac{p}{3}p_k^{e_i^x} + (1-p)(1-p_k^{e_i^x}),
\end{aligned}
\end{equation}
where we have used the fact that in a depolarizing channel $P(e^z=0,e^x=0)=1-p$ and $P(e^z=1,e^x=0)=P(e^z=0,e^x=1)=P(e^z=1,e^x=1)=\frac{p}{3}$. $p_k^{e_i^x}=P(e_{i,k}^x=1)$ is calculated in the $X$-decoder including all the messages proceeding from the nodes below the corresponding node $e_{i,k}^x$ but excluding the prior probability, $P(e^x=1)=2p/3$. Then, the value of $p_k^{e_{\hat{i}}^z}$ obtained in (\ref{eqn:p_corr}) is used in the
$Z$-decoder as  {\it a priori} probability of a $Y$- or $Z$-error in the
corresponding qubit in lieu of the prior $P(e_k^z=1)=2p/3$, which is no longer utilized. 

Denoting by $L_k^{e_i^x}=\log\frac{1-p_k^{e_i^x}}{p_k^{e_i^x}}$, we can easily obtain $L_k^{e_{\hat{i}}^z}=\log\frac{1-p_k^{e_{\hat{i}}^z}}{p_k^{e_{\hat{i}}^z}}$ as a function of $L_k^{e_i^x}$
\begin{equation}
\label{eqn:L_ez}
L_k^{e_{\hat{i}}^z}=
\log\frac{1+3e^{{L_k^{e_i^x}}+L^p}}{1+e^{{L_k^{e_i^x}}}}, \text{ with } L^p=\log\frac{1-p}{p}.
\end{equation}

The message passed from the $Z$-decoder to the $X$-decoder is
obtained by substituting x by z (and vice versa) in (\ref{eqn:L_ez}).



\section{Discrete Density Evolution}\label{sec:DDE}
We now implement Discrete Density Evolution, DDE, originally proposed in \cite{dde} for classical codes, to predict the performance of the proposed LDGM-based quantum codes. Although DDE is a standard tool commonly used in classical coding, to our knowledge it has never been discussed in the quantum coding literature. 

\subsection{DDE in Classical vs Quantum LDPC/LDGM Codes}
Let us first consider a classical regular LDGM code, with its associated graph representing equation $c=G_c^Tu$, where $G_c$ is the non-systematic section of the generator matrix. Input bit nodes
$u=(u_1,u_2,\dots,u_K)^T$ have degree $d_v$ and parity bit nodes
$c=(c_1,c_2,\dots,c_M)^T$ have degree $d_c$. In addition to prior knowledge about $u$, the noisy observations of $c$ (denoted as $c'$), and---for systematic codes---also the noisy observations of $u$ (denoted as $u'$), are available for decoding\footnote{The LDGM code may be used as either a channel or a source code, depending on whether the output bits contain the input bits in addition to the parity bits, as well as on the ratio between $K$ and $M$.}.

Let $v=\log\frac{P(u=0)}{P(u=1)}$ be the LLR message
transferred from a generic input bit node to an associated parity
bit node, and $w$ the LLR message from a generic parity bit node to an associated input bit node. DDE follows the evolution
of the discretized version of the probability density functions (referred to as probability mass functions in the sequel) of $u$ and $v$ as the messages are exchanged in the associated graph. Notice that for a classical LDPC code, which can be represented by a graph associated with equation $0=H_pc$, where $H_p$ is now the parity-check matrix and $c$ is a codeword, DDE can be seen as a particular case of DDE for LDGM codes, since the equation for LDPC codes is a particular case of that of LDGM codes when the left-hand side of $c=G_c^Tu$ is $0$, and in the right-hand side $u=c$ and $G_c^T=H_p$. Thus, next section presents DDE for the most general case of classical LDGM codes, and in the sequel the DDE equations will be particularized for the specific graph section under consideration. 

It is important to remark that, for classical LDPC codes, the graph mentioned in the previous paragraph (associated with equation $0=H_pc$, where $c$ is a codeword and $c'=c\oplus e$), is different from the one usually used in message passing, which corresponds to equation $s=H_pe$, where $s$ is the syndrome vector and $e$ is the error pattern vector. However, both equations are equivalent. In the quantum case, the latter equation has perfect physical meaning and is associated with (\ref{d},\ref{s}) and the graph depicted in Fig. \ref{fig:decoding} by considering $e=\left(%
\begin{array}{c}
  e_1 \\
  e_2 \\
\end{array}%
\right)$ and $H_p=M_1 [P^T \; I ]$. The syndrome $s$ is obtained through quantum measurements and $e=\left(%
\begin{array}{c}
  e_1 \\
  e_2 \\
\end{array}%
\right)$ corresponds to the binary representation of the quantum error operators. 

Because of the aforementioned equivalence between the two equations in the classical domain, even though $0=H_pc$, with $c'=c\oplus e$, has no physical meaning in the quantum domain, the error predictions obtained by applying DDE on its corresponding graph  will also predict the error probability when message passing is performed on the graph associated with  $s=H_pe$, which is the one we utilize to decode our proposed quantum codes (see Fig. \ref{fig:decoding}). The reason to perform DDE on the graph corresponding to $0=H_pc$, where as indicated before $H_p=M_1 [P^T \; I ]$ (i.e., the same graph as in Fig. \ref{fig:decoding} but with all syndrome nodes equal to 0 and the bottom nodes being the transmitted classical codeword rather than the error pattern), is that, assuming that the all-zero sequence is transmitted, we can easily predict the node error probability from the  probability mass functions of the exchanged messages.

\subsection{DDE for Classical LDGM Codes}

Coming back to message passing for classical LDGM codes, the LLR message $v$ departing from an input bit node in the graph defined by equation $c=G_c^Tu$ is updated as \cite{dde}
\begin{flalign}
&v = \sum_{i=0}^{d_v-1} w_{i} \label{eq:LLRvLDPC},
\end{flalign}
where $w_0$ is the prior LLR message associated with the corresponding input bit node, and $w_i$, $i=1,2,\dots,d_v-1$ are the LLR
messages proceeding from the associated parity bit nodes except for the one to which $v$ is sent. The probability mass function of $v$ is given
as
\begin{flalign}
&P(v) =
P(w_0)\bigotimes\left(\bigotimes_{i=1}^{d_v-1}P(w_i)\right)\label{eq:pmfvLDPC},
\end{flalign}
where $\bigotimes$ denotes the convolution operation.

The LLR message exchanged from a parity
bit node to an input bit node, $w$, is updated by the ``tanh rule" in
\cite{Hagenauer} and obtained as
\begin{flalign}
&\tanh(\frac{w}{2}) = \prod_{j=0}^{d_c-1}\tanh(\frac{v_{j}}{2}),
\label{eq:tanhuLDGM}
\end{flalign}
where $v_j$, $j=1,2,\dots,d_c-1$ are the LLR messages proceeding from the
incident input bit nodes except for the message from
the bit node where $w$ is sent to, and $v_0$ is the prior LLR 
message corresponding to the parity bit node. Expression
(\ref{eq:tanhuLDGM}) can be calculated iteratively as
\begin{flalign}
&w = \mathrm{R}(v_0,
\mathrm{R}(v_1,\mathrm{R}(v_2,\dots,\mathrm{R}(v_{d_c-2},v_{d_c-1})))),
\label{eq:LLRuLDGM}
\end{flalign}
where the function $R$ is defined as
\begin{flalign}
\mathrm{R}(a,b)=\mathrm{Q}(2\tanh^{-1}(\tanh\frac{a}{2}\tanh\frac{b}{2})),\label{eq:R}
\end{flalign}
and $\mathrm{Q}$ is a quantization function. The probability mass function of $\mathrm{R}(a,b)$ is denoted as $R(P_a,P_b)$ and obtained as
\begin{flalign}
R(P_a,P_b)=P_c[k]=\sum_{(i,j):k\delta=\mathrm{R}(i\delta,j\delta)}P_a[i]P_b[j],\label{eq:pmfR}
\end{flalign}
where $\delta$ is the quantization step. Based on
(\ref{eq:LLRuLDGM}), (\ref{eq:R}) and (\ref{eq:pmfR}), the
probability mass function of the LLR message $w$, $P(w)$ can be
calculated in an iterative way.

The {\it a posteriori} LLR message of an input bit node $u$, used to make a decision, is calculated
as
\begin{flalign}
&\mathrm{LLR}(u) = \sum_{i=0}^{d_v} w_{i} \label{eq:LLReLDGM},
\end{flalign}
and the probability mass function of $\mathrm{LLR}(u)$ is given by
\begin{flalign}
&P(\mathrm{LLR}(u)) =
P(w_0)\bigotimes\left(\bigotimes_{i=1}^{d_v}P(w_i)\right)\label{eq:pmfeLDGM}.
\end{flalign}

Assuming that the all zero sequence is transmitted, the
probability of error for an input bit node is obtained as
\begin{flalign}
&P_{e}=\int_{-\infty}^{0}P\big(\mathrm{LLR}(u)\big)du.\label{eq:PeLDGM}
\end{flalign}

\subsection{DDE for the Proposed CSS Codes}\label{sec:DDE_struc1}

We now  discuss DDE for the proposed CSS codes described in Figs. \ref{fig:decoding} and \ref{figure1}\footnote{A very similar development can be performed for the fault-tolerant quantum codes proposed in Section IV.}. We denote by 
 $m_{ch}$ the prior LLR message (depending on the channel), with probability mass function $p_{m_{ch}}$. $m(r,t)$ denotes the random
variable representing the quantized LLR message exchanged from a
node of class $r$ to a node of class $t$ ($r$, $t$ $\in\{e_1, e_2,
c, d, s\}$). The probability mass function of $m(r,t)$ is denoted as
$p_{m(r,t)}$. Extending the DDE technique in \cite{dde} to the graphs
presented in Figs. \ref{fig:decoding} and \ref{figure1}, we can obtain the equations
for the updating of the probability mass functions of the quantized
LLR messages. As described in Section II.E, the lower part of the graph in Fig. \ref{fig:decoding}, where information is exchanged between nodes $e_1$ and $c$, and between nodes $e_2$/$c$ and $d$ can be seen as a classical non-systematic LDGM code, while the upper layer in Fig. \ref{fig:decoding} can be seen as a classical LDPC code. The proper equations to be used in each section are specified in the sequel. From the final probability mass functions of the LLRs of $e_1$ and $e_2$, we can easily predict the probability of error. Recall that in the proposed codes matrix $P$ is regular with size $n_1 \times n_1$, $n_1=N/2$, and degrees $(y,y)$. 

\subsubsection{\bf Lower Layer: Information Passing between Nodes $e_1$ and $c$}
The quantized LLR message exchanged from nodes $e_1$ to nodes $c$, denoted as $m(e_1,
c)$, is the sum of the quantized LLR messages $m(c,e_1)$ exchanged from nodes $c$ to node $e_1$ (except for node $c$ where $m(e_1,c)$ is sent to), and the
prior LLR message (depending on the channel), denoted as $m_{ch}$. The
probability mass function of $m(e_1, c)$ is calculated as the discrete convolution of the probability
mass functions of $m(c, e_1)$ and the probability mass function of
$m_{ch}$ as
\begin{equation}
p_{m(e_1, c)}=p_{m_{ch}}\bigotimes \left(\bigotimes ^{y-1}
p_{m(c, e_1)}\right) \label{eq:pmf_ve1c},
\end{equation}
where $p_{m_{ch}}$ denotes the probability mass function of the prior LLR message and is calculated as indicated in (\ref{exchange}).


The quantized LLR message exchanged from a
node $c$ to a node $e_1$, $m(c, e_1)$, is obtained by
utilizing the ``tanh rule'' in \cite{Hagenauer}, and the
corresponding probability mass function is
\begin{equation}
p_{m(c, e_1)}=R\left(p_{m(d, c)}, R ^{y-1} p_{m(e_1, c)}\right)
\label{eq:pmf_ue1c},
\end{equation}
where $p_{m(d, c)}$ is the probability mass function of the
quantized LLR messages exchanged from nodes $d$ to nodes $c$, calculated in (\ref{eq:pmf_ucd}), and $R^{y-1} p_{m(e_1, c)}$
is the probability mass function of the LLR message
$R^{y-1}(m(e_1,c))$, calculated by applying (\ref{eq:LLRuLDGM}) to $v_1=v_2=\dots =v_{y-1}=m(e_1,c)$.
$R ^{y-1}p_{m(e_1, c)}$ is calculated by iteratively applying (\ref{eq:R}) and
(\ref{eq:pmfR}).

\subsubsection{\bf Lower Layer: Information Passing between Nodes $e_2$/$c$ and $d$}

As shown in Fig. \ref{fig:decoding}, there is only one connection
between nodes $e_2$ and nodes $d$. Thus, the quantized LLR message
exchanged from a node $e_2$ to a node $d$, denoted as $m(e_2, d)$, is just 
the prior LLR message,
$m_{ch}$. Then, the probability mass function of $m(e_2, d)$ is the
same as the probability mass function of $m_{ch}$
\begin{equation}
p_{m(e_2, d)}=p_{m_{ch}} \label{eq:pmf_ve2d}, 
\end{equation}
where $p_{m_{ch}}$ is calculated as indicated in (\ref{exchange}).

The quantized LLR message exchanged from a node $c$ to a node $d$ is
denoted as $m(c, d)$. We obtain it by utilizing the ``tanh rule'', and
calculate its probability mass function as
\begin{equation}
p_{m(c, d)}=R ^{y} p_{m(e_1, c)}, \label{eq:pmf_vcd}
\end{equation}
where $R^{y} p_{m(e_1, c)}$
is the probability mass function of the LLR message
$R^{y}(m(e_1,c))$, calculated by applying (\ref{eq:LLRuLDGM}) to $v_1=v_2= \dots =v_{y}=m(e_1,c)$.
$R ^{y}p_{m(e_1, c)}$ is obtained by iteratively applying (\ref{eq:R}) and
(\ref{eq:pmfR}).

In the middle layer shown in Fig. \ref{fig:decoding}, the quantized
LLR message exchanged from a node $d$ to a node $c$ is denoted as
$m(d,c)$, and the quantized LLR message exchanged from a node $d$ to
a node $e_2$ is denoted as $m(d,e_2)$. The quantized LLR message
$m(d,c)$ is obtained by utilizing the ``tanh rule'', and
contains the information of the quantized LLR message $m(e_2, d)$
and the quantized LLR message $m(d)$, which as explained in the sequel is the sum of all the
incoming quantized LLR messages to a randomly chosen node $d$ proceeding from nodes $s$. Specifically,
\begin{equation}
p_{m(d, c)}=R\left( p_{m(d)}, p_{m(e_2, d)}\right),
\label{eq:pmf_ucd}
\end{equation}
where $p_{m(d)}$ is calculated as indicated in (\ref{eq:pmf_db0}).

Similarly, the probability mass function of the LLR messages exchanged from nodes $d$ to nodes $e_2$, $m(d, e_2)$, can be calculated as
\begin{equation}
p_{m(d, e_2)}=R\left(p_{m(d)}, p_{m(c, d)}\right)
\label{eq:pmf_ue2d}.
\end{equation}

\subsubsection{\bf Upper Layer: Information Passing between Nodes $d$ and $s$, and calculation of $p_{m(d)}$}

In matrix $M$, nodes $d$ are considered as parity bit nodes
and nodes $s$ are syndrome nodes. 
From Fig.~\ref{figure1}, we observe that each node $s_A$ is connected to a single node $d_A$, and vice versa. Each node $s_B$ has degree two, with both edges connected exclusively to nodes $d_B$, while for each node $d_B$ one edge connects to a node $s_B$ and the remaining $w_d-1$ edges connect to nodes $s_C$. Nodes $d_C$ are connected exclusively to nodes $s_C$, while nodes $s_C$ may be connected to nodes $d_B$ and $d_C$.

The probability mass function of $m(s_A, d_A)$, the LLR message exchanged from a syndrome node
$s_A$ to a node $d_A$, is
\begin{equation}
p_{m(s_A, d_A)}=p_{m(s_A)}=p_{m(s_0)},
\label{eq:pmf_vdasA}
\end{equation}
where $p_{m(s_A)}$ denotes the probability mass function of the LLR of a syndrome node $s_A$. Since the syndromes are all zero (a codeword is to be recovered), $p_{m(s_A)}=p_{m(s_0)}$, where $p_{m(s_0)}$  is a delta at infinity.

Syndrome nodes $s_B$ connect with two nodes $d_B$, and also satisfy $p_{m(s_B)}=p_{m(s_0)}$. Therefore, the probability mass function of $m(s_B, d_B)$, the quantized LLR messages exchanged from nodes
$s_B$ to nodes $d_B$, is
\begin{equation}
p_{m(s_B, d_{B})} = p_{m(d_{B},s_B)}, 
\label{eq:pmf_udBsB}
\end{equation}
where $p_{m(d_{B},s_B)}$ is calculated in (\ref{eq:pmf_vdBsB}).


Syndrome nodes $s_C$ connect with nodes $d_B$ and $d_C$. Assuming that the connections between the nodes are randomly set, and that the probability mass functions of the LLR messages $m(s_C,d_B)$ and $m(s_C,d_C)$ are the same ($m(s_C,d_B)=m(s_C,d_C)=m(s_C,d)$), which are reasonable assumptions for the set of parameters under consideration, the probability mass function of the LLR messages departing from a node $s_C$, $m(s_C,d)$, can be calculated by applying the ``tanh rule'' as

\begin{equation}
p_{m(s_C, d)}=R ^{w_s-1}
p_{m(d,s_C)}, \label{eq:pmf_udsC}
\end{equation}
where $w_s$ is the degree of the nodes $s_C$\footnote{In general, $w_s$ is not an integer: some nodes $s_C$ will have degree $\lfloor w_s \rfloor$, while the degree for the other nodes $s_C$ will be $\lceil w_s \rceil$. In such cases, $p_{m(s_C, d)}=\alpha_{\lfloor} R^{\lfloor w_s \rfloor -1} p_{m(d,s_C)}+\alpha_{\lceil} R^{\lceil w_s \rceil -1} p_{m(d,s_C)}$, where $\alpha_{\lfloor}$ and $\alpha_{\lceil}$ are the fraction of edges departing nodes $s_C$ with degrees $\lfloor w_s \rfloor$ and $\lceil w_s \rceil$, respectively.} and $p_{m(d,s_C)}$ is the probability mass function of an LLR message (chosen at random) arriving at a node $s_C$ from a node $d$. It can be calculated as 
\begin{equation}
p_{m(d,s_C)} = \alpha_{d_B} p_{m(d_B, s_{C})}+\alpha_{d_C} p_{m(d_C, s_{C})}, \label{eq:pmf_udsD}
\end{equation}
where $p_{m(d_B, s_{C})}$ and $p_{m(d_C, s_{C})}$ are calculated in (\ref{eq:pmf_vdBsC}) and (\ref{eq:pmf_vdbsB}), respectively, and $\alpha_{d_B}$ and $\alpha_{d_C}$ represent the fraction of edges arriving at a node $s_C$ from nodes $d_B$ and $d_C$, respectively. Specifically
\begin{equation*}
\alpha_{d_B} = \frac{2N_b(w_d-1)}{2N_b(w_d-1)+(\frac{N}{2} - N_a-2N_b)w_d}, 
\qquad
\alpha_{d_C} = 1 - \alpha_{d_B}.
\end{equation*}

There are two types of LLR messages departing from a node $d_B$ towards a syndrome node: i) the messages  to nodes $s_B$, $m(d_B,s_B)$, which combine $w_d-1$ incoming messages from nodes $s_C$ and the messages received in node $d_B$ from nodes $c$ and $e_2$, and ii) the messages to nodes $s_C$, which combine one incoming message from a node $s_B$, $w_d-2$ incoming messages from nodes $s_C$, and the messages received in node $d_B$ from nodes $c$ and $e_2$. The probability mass functions of these messages are
\begin{equation}
p_{m(d_{B}, s_{B})}=R\left(p_{m(c, d)},p_{m(e_2, d)}\right)\bigotimes ^{w_d-1} p_{m(s_C, d)},
\label{eq:pmf_vdBsB}
\end{equation}

\begin{equation}
p_{m(d_{B}, s_{C})}=R\left(p_{m(c, d)},p_{m(e_2, d)}\right)\bigotimes p_{m(s_B, d_{B})}
\bigotimes ^{w_d-2} p_{m(s_C, d)},
\label{eq:pmf_vdBsC}
\end{equation}
where $p_{m(e_2, d)}$, $p_{m(c, d)}$, $p_{m(s_B, d_{B})}$ and $p_{m(s_C, d)}$ are calculated in (\ref{eq:pmf_ve2d}), (\ref{eq:pmf_vcd}), (\ref{eq:pmf_udBsB}) and (\ref{eq:pmf_udsC}), respectively.


The quantized LLR messages exchanged from nodes $d_C$ to nodes
$s_C$, denoted as $m(d_C, s_{C})$, consist of all the quantized
LLR messages proceeding from connected nodes $s_C$ except
from the one where $m(d_C, s_{C})$ is sent to, and the LLR
messages received in node $d_C$ from nodes $c$ and $e_2$. Thus, the probability mass
function of $m(d_C, s_{C})$ is obtained as
\begin{equation}
p_{m(d_C, s_{C})}=R\left(p_{m(c, d)},p_{m(e_2, d)}\right) \bigotimes ^{w_d-1} p_{m(s_C, d)},
\label{eq:pmf_vdbsB}
\end{equation}
where $p_{m(e_2, d)}$, $p_{m(c, d)}$, and $p_{m(s_C, d)}$ are calculated in (\ref{eq:pmf_ve2d}), (\ref{eq:pmf_vcd}), and (\ref{eq:pmf_udsC}), respectively.

As described when discussing the DDE in the lower layer, the quantized LLR message $m(d)$, consisting of the sum of all the incoming quantized LLR messages arriving at a randomly chosen node $d$ from connected nodes $s$, is required to obtain the probability mass functions of $m(d,c)$ and $m(d,e_2)$ in (\ref{eq:pmf_ucd}) and (\ref{eq:pmf_ue2d}), respectively. Since nodes $d$ are partitioned into the three categories $d_A$, $d_B$, and $d_C$, the probability mass function of $m(d)$ is calculated as the weighted sum of the respective probability mass functions, $p_{m(d_A)}$, $p_{m(d_B)}$, and $p_{m(d_C)}$. 

Taking into account that a node $d_A$ only receives an incoming message from a node $s_A$, a node $d_B$ receives one message from a node $s_B$ and $w_d-1$ messages from nodes $s_C$, and a node $d_C$ receives all its $w_d$ messages from nodes $s_C$, the corresponding probability mass functions are 
\begin{equation}
p_{m(d_A)} = p_{m(s_A, d_A)} = p_{m(s_0)},
\label{eq:pmf_da0}
\end{equation}
where $p_{m(s_0)}$  is a delta at infinity, independent of the iteration number.

\begin{equation}
p_{m(d_B)} =  
\bigotimes^{w_d - 1} p_{m(s_C, d)} \bigotimes p_{m(s_B, d_B)},
\label{eq:pmf_db}
\end{equation}
where $p_{m(s_B, d_B)}$ and $p_{m(s_C, d)}$ are calculated in (\ref{eq:pmf_udBsB}) and (\ref{eq:pmf_udsC}), respectively.

\begin{equation}
p_{m(d_C)} = 
\bigotimes^{w_d} p_{m(s_C, d)}.
\label{eq:pmf_dc}
\end{equation}

Finally, for a randomly chosen node $d$, we obtain the probability mass function of $m(d)$, to be used in (\ref{eq:pmf_ucd}) and (\ref{eq:pmf_ue2d}), as
\begin{equation}
p_{m(d)} =
\frac{N_a}{N/2} \, p_{m(d_A)} +
\frac{2N_b}{N/2} \, p_{m(d_B)} +
\frac{N/2 - N_a - 2N_b}{N/2} \, p_{m(d_C)}.
\label{eq:pmf_db0}
\end{equation}

\subsubsection{\bf Predicted Error}
Based on the final probability mass functions of the quantized LLR
messages, after a sufficient number of iterations have been performed, we calculate the sum of all the quantized LLR messages into a
given node $e_1$ and $e_2$, which are denoted as $m(e_1)$ for
$e_1$, and $m(e_2)$ for $e_2$. Since  we are transmitting the all-zero sequence, the probability of error for a given
node $e_1$, $P_e(e_1)$, or $e_2$, $P_e(e_2)$, can be calculated by integrating the discretized probability density function of $m(e_1)$ or $m(e_2)$, the final LLRs of $e_1$ and $e_2$, from
negative infinity to zero

\begin{equation}
p_{m(e_1)}=p_{m_{ch}}\bigotimes p_{m(e_1)}^{ext}, \text{where }p_{m(e_1)}^{ext}=\left(\bigotimes ^y p_{m(c,
e_1)}\right) \label{eq:pmf_e1}
\end{equation}

\begin{equation}
P_e(e_1)=\int_{-\infty}^{0} p_{m(e_1)}(x) dx\label{eq:Pe_e1}
\end{equation}

\begin{equation}
p_{m(e_2)}=p_{m_{ch}}\bigotimes p_{m(e_2)}^{ext}, \text{where }p_{m(e_2)}^{ext}=p_{m(d, e_2)}\label{eq:pmf_e2}
\end{equation}

\begin{equation}
P_e(e_2)=\int_{-\infty}^{0} p_{m(e_2)}(x) dx\label{eq:Pe_e2}.
\end{equation}

\subsubsection{\bf {Exploiting the Correlation between $e^x$ and $e^z$ Errors}}

The development above does not exploit the correlation between the $e^x$ and $e^z$ errors introduced by the depolarizing channel. Indeed, up to now DDE has been performed independently for the $X$- and the $Z$-decoders, and in both decoders the probability mass function for the prior LLR message $m_{ch}$ that depends on the depolarizing channel, $p_{m_{ch}}$, used in (\ref{eq:pmf_ve1c}, \ref{eq:pmf_ve2d}, \ref{eq:pmf_e1}, \ref{eq:pmf_e2}), has been calculated based on the marginals of the depolarizing channel. Specifically, recall from Section VI.A that in DDE  we are using the notation $e=\left( \begin{array}{c}
  e_1 \\
  e_2 \\
\end{array} \right)$  to represent a codeword rather than an error vector, so that for the $X$-decoder\footnote{The same development holds for the $Z$-encoder by exchanging x and z.} $e_i(x)'=e_i(x)\oplus e_i^x, i\in\{1,2\}$, where $e_i(x)$ is a bit of the codeword, $e_i(x)'$ is the corrupted version, and $e_{i}^x=1$ represents an X or a Y error. Then, since we are transmitting the all-zero sequence ($e_i(x)=0$), the probability mass of the observed, noisy, bit $e_i(x)'$ is $P(e_i(x)'=1)=1-P(e_i(x)'=0)=p'$, where $p'=\frac{2p}{3}$ and $p$ is the depolarizing parameter. Thus,  random variable $m_{ch}(x)$, where $(x)$ indicates that we are using the X-decoder, is

\begin{equation}
m_{ch}(x)=\log\frac{P(e_i(x)=0)}{P(e_i(x)=1)}=\left\{
\begin{array}{ll}
     \log\frac{P(e_{i}^x=0)}{P(e_{i}^x=1)}=\log\frac{1-p'}{p'}& \text{with prob. $1-p'$ (when }  e_i(x)'= 0) \\
    \log\frac{P(e_{i}^x=1)}{P(e_{i}^x=0)}=\log\frac{p'}{1-p'}& \text{with prob. $p'$ (when }  e_i(x)'= 1)
\end{array}.
\right.
\label{exchange}
\end{equation}

In order to exploit the correlation between the $e^x$ and $e^z$ errors in DDE, we need to consider the relationship between $L^{e_{\hat{i}}^z}_k$ and $L^{e_i^x}_k$ as defined in (\ref{eqn:L_ez}), so that rather than having independent decoders for the graphs associated with the $H$ and $G$ matrices, information from the $X$-decoder will be transferred to the $Z$-decoder and vice versa. 

Denoting by $m(e_1(x))^{ext}$ and $m(e_2(x))^{ext}$ the random variables of the extrinsic LLRs of $e_1$ and $e_2$, defined in the $X$-decoder by probability masses $p_{m(e_1(x))}^{ext}$ and $p_{m(e_2(x))}^{ext}$, calculated using (\ref{eq:pmf_e1}) and (\ref{eq:pmf_e2}), respectively, and taking into account that $e_i(x)'=e_i(x)\oplus e_{i}^x$, we obtain random variable $L^{e_{i}^x}$ as

\begin{equation}
\label{eqn:e(x)}
L^{e_{i}^x}=\log\frac{P(e_i^x=0)}{P(e_i^x=1)}=\left\{
\begin{array}{ll}
     \log\frac{P(e_i(x)=0)}{P(e_i(x)=1)}=m(e_i(x))^{ext}& \text{if } e_i(x)'=0 \\
     \log\frac{P(e_i(x)=1)}{P(e_i(x)=0)}=-m(e_i(x))^{ext}& \text{if } e_i(x)'=1
\end{array}.
\right.
\end{equation}

Using (\ref{eqn:L_ez}), we obtain
\begin{equation}
\label{eqn:L_ez(x)}
L^{e^z_{\hat{i}}}=f(L^{e_{i}^x},L^p)=
\log\frac{1+3e^{{L^{e_{i}^x}}+L^p}}{1+e^{{L^{e_{i}^x}}}}, \text{ where } L^p=\log\frac{1-p}{p}.
\end{equation}

Recall from (\ref{eqn:L_ez}) that in the $Z$-decoder $L_k^{e_{\hat{i}}^z}$ was used as prior knowledge in lieu of the marginal obtained from the depolarizing channel. However, in DDE we are considering the evolution of the probability mass of the LLRs of the bits $e_{\hat{i}}(z)$, which are related to the error $e_{\hat{i}}^z$ by $e_{\hat{i}}(z)'=e_{\hat{i}}(z)\oplus e^{z}_{\hat{i}}$. The ``prior" to use in the $Z$-decoder, proceeding from the $X$-decoder and the depolarizing channel, is obtained as  

\begin{equation}
\label{eqn:e(z)}
m_{ch}^{\hat{i}}(z)=L^{e_{\hat{i}}(z)}=\log\frac{P(e_{\hat{i}}(z)=0)}{P(e_{\hat{i}}(z)=1)}=\left\{
\begin{array}{ll}
     \log\frac{P(e_{\hat{i}}^z=0)}{P(e_{\hat{i}}^z=1)}=L^{e_{\hat{i}}^z}& \text{if } e_{\hat{i}}(z)'=0 \\
    \log\frac{P(e_{\hat{i}}^z=1)}{P(e_{\hat{i}}^z=0)}=-L^{e_{\hat{i}}^z}& \text{if } e_{\hat{i}}(z)'=1
\end{array}
\right.
\end{equation}

Combining (\ref{eqn:e(x)},\ref{eqn:L_ez(x)},\ref{eqn:e(z)}), and since we are transmitting the all zero sequence, $e_i(x)=e_{\hat{i}}(z)=0$, we obtain

\begin{equation}
m_{ch}^{\hat{i}}(z)=\left\{
\begin{array}{ll}
     f(m(e_i(x))^{ext},L^p)& \text{with prob. $1-p$  } \text{(when } e_i(x)'=0, e_{\hat{i}}(z)'=0) \\
     f(-m(e_i(x))^{ext},L^p)& \text{with prob. $p/3$  } \text{(when } e_i(x)'=1, e_{\hat{i}}(z)'=0) \\
     -f(m(e_i(x))^{ext},L^p)& \text{with prob. $p/3$  } \text{(when } e_i(x)'=0, e_{\hat{i}}(z)'=1) \\
     -f(-m(e_i(x))^{ext},L^p)& \text{with prob. $p/3$  } \text{(when } e_i(x)'=1, e_{\hat{i}}(z)'=1),  
\end{array}
\right.
\label{eq:pass}
\end{equation}
where function $f$ is defined in (\ref{eqn:L_ez(x)}). Notice that we have defined random variable $m_{ch}^{\hat{i}}(z)$
as a function of random variable $m(e_i(x))^{ext}$, for which we know its probability mass function, $p_{m(e_i(x))}^{ext}$, from (\ref{eq:pmf_e1}) and (\ref{eq:pmf_e2}). In order to obtain the probability mass function of $m_{ch}^{\hat{i}}(z)$, which will be used in (\ref{eq:pmf_ve1c}) and (\ref{eq:pmf_ve2d}) to continue with the DDE process in the $Z$-decoder, we can simply generate Monte Carlo samples of $m(e_i(x))^{ext}$ using $p_{m(e_i(x))}^{ext}$ and obtain the corresponding samples of $m_{ch}^{\hat{i}}(z)$ by applying (\ref{eq:pass}). Then, we can approximate the probability mass of $m_{ch}^{\hat{i}}(z)$ by its histogram. Information passing from the $Z$-decoder to the $X$-decoder is performed similarly substituting x by z (and vice versa).

\section{Simulation Results}
In all results in this section, a depolarizing channel of parameter $p$ is considered. For comparison purposes, we consider quantum codes
with the same rate, $R_Q$, and block length, $N$, as in \cite{mackay} ($R_Q \cong
1/4$ and $N=19014$), encoding $K=4752$ qubits into
$N=19014$ qubits. Matrix $P$ corresponds to a $1/2$ LDGM code and has size
$9507 \times 9507$, so that both $\tilde{G}$ and $\tilde{H}$, the generator and parity-check matrices of a regular LDGM code of degrees $(y,y)$, have
size $9507 \times 19014$. Matrix $M$, with size $7131 \times 9507$,
is used to transform $\tilde{H}$ and $\tilde{G}$ into $H$ and $G$. The specific form of matrix $M$ will depend on the utilized structure. Both the original structure proposed in Fig. \ref{figure1} and the fault-tolerant codes described in Fig. \ref{figure3} will be investigated. 

\begin{figure}[!t]
\centering
\hspace{-1cm}
\begin{minipage}{\textwidth}
\centering
\subfloat{\label{fig:fig1}\includegraphics[width=0.46\textwidth]{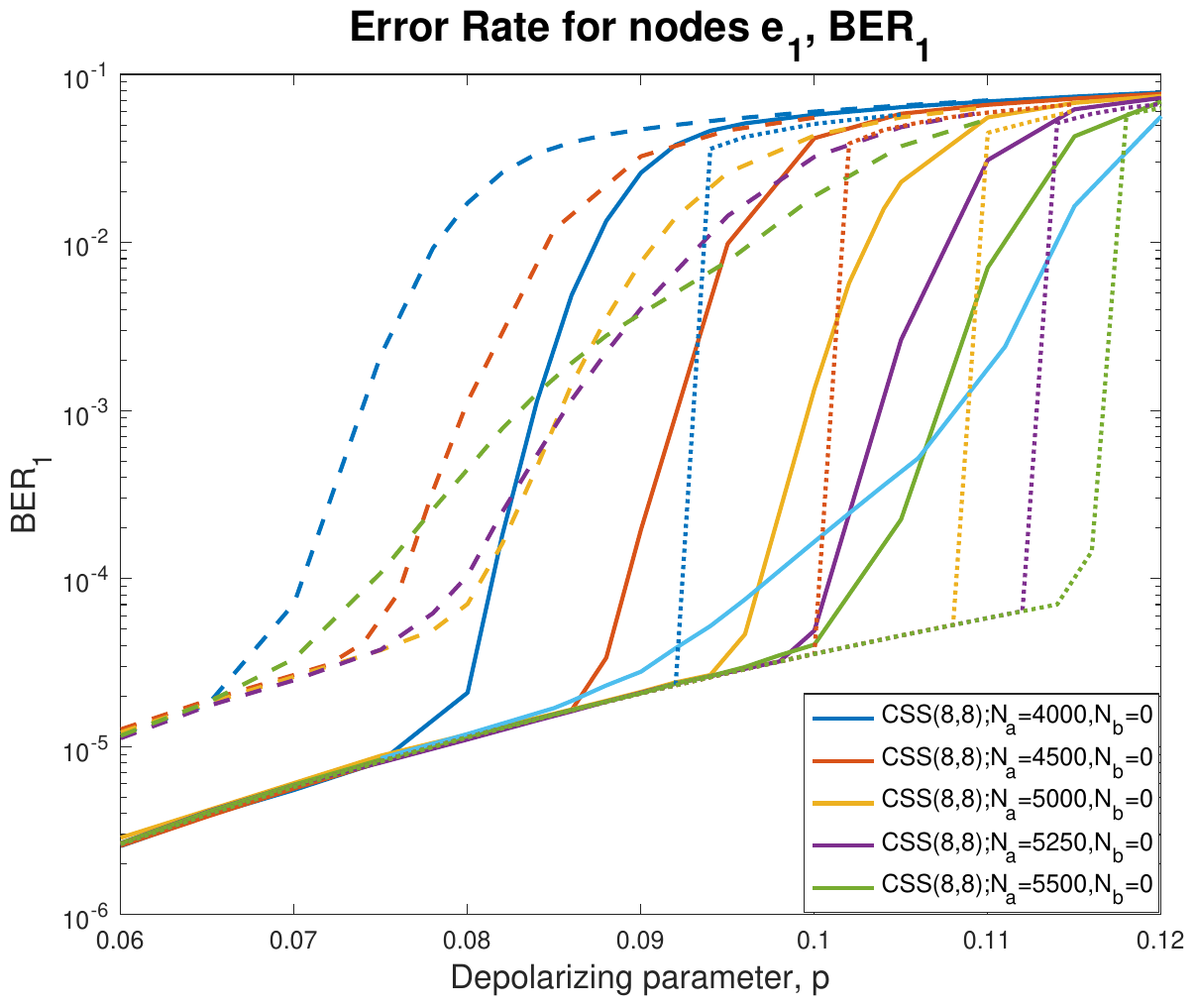}}
\hfill
\subfloat{\label{fig:fig2}\includegraphics[width=0.46\textwidth]{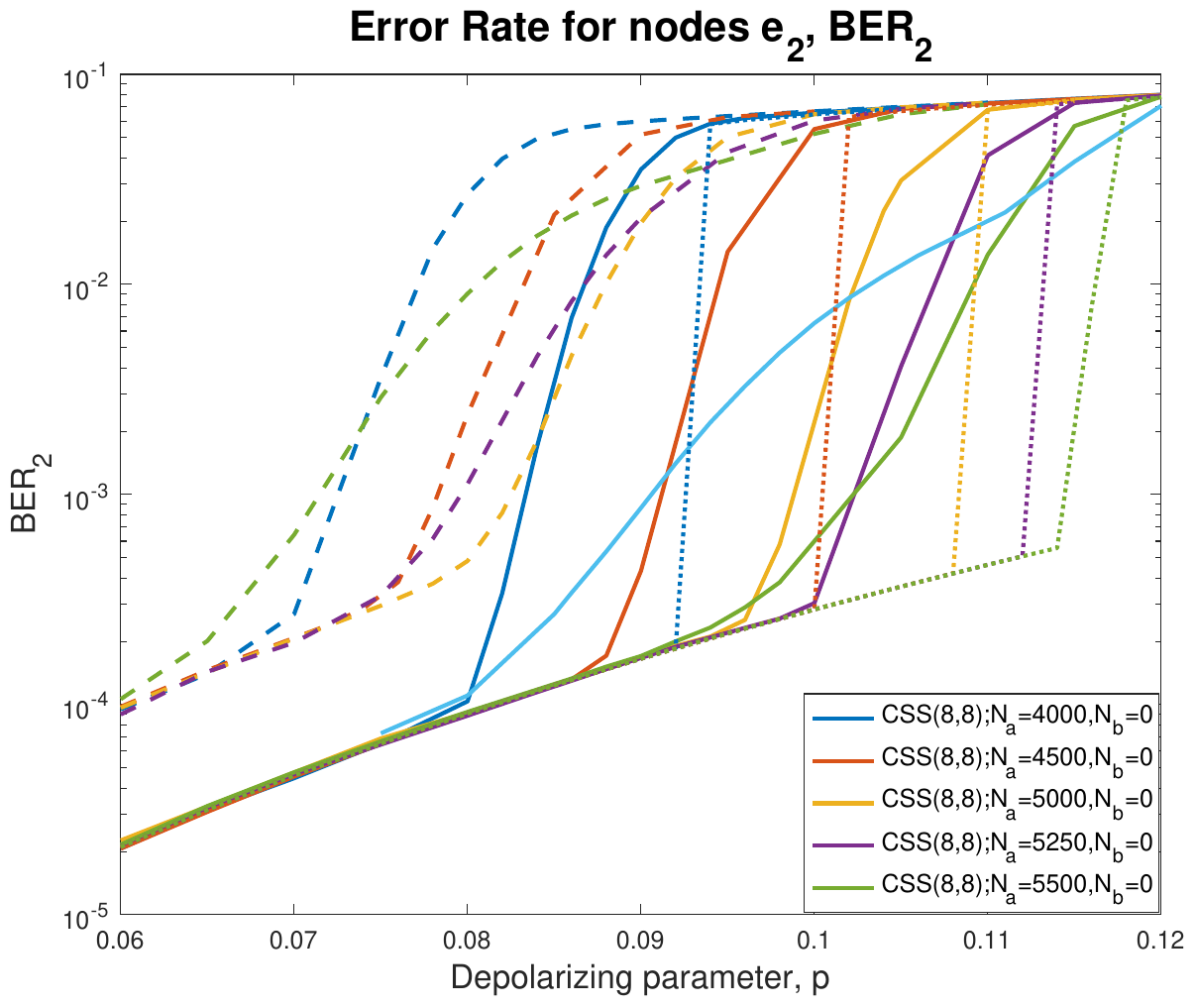}}
\end{minipage}
\caption{DDE predictions and simulated residual error rate in both $e_1$ and $e_2$ for a family of quantum codes designed using the original structure proposed in Fig. \ref{figure1} with an LDGM code of degrees $(8,8)$ in the bottom layer, $N_b=0$ and different values of $N_a$. A depolarizing channel is considered. The rightmost curves correspond to the case where the correlation between the $e^x$ and $e^z$ errors is exploited in the decoder via information exchange between the X and Z sections of the graph (continuous lines for simulation results and dotted lines for DDE predictions). The leftmost curves (dashed lines) correspond to the simulation results when this exchange does not occur.}
\vspace{-0.8cm}
\label{fig:NP_sA8,8}
\end{figure}
\subsection{Original Structure}
We first consider the original structure proposed in Fig. \ref{figure1}. In all cases, the degree of the nodes $d$ not connected to nodes $s_A$ is fixed to $w_d=3$, as this value resulted in the best performance (results not shown due to space constraints). Fig. \ref{fig:NP_sA8,8}  compares the DDE predictions with the simulation results for LDGM codes with degree $(8,8)$ when $N_b=0$ and different values of $N_a$ are considered. Specifically, we plot the residual error rate (BER) in both $e_1$ and $e_2$ versus the depolarizing parameter, $p$. The rightmost curves correspond to the case where the correlation between the $e^x$ and $e^z$ errors is exploited in the decoder via information exchange between the X and Z sections of the graph (continuous lines for simulation results and dotted lines for DDE predictions), while the leftmost curves correspond to the case in which this exchange does not occur (dashed lines). It is remarkable that exploiting the correlation between the $e^x$ and $e^z$ errors (rightmost curves in the figure) leads to significant improvements in performance, both in terms of convergence thresholds and error floors. This occurs for all the range of $N_a$, but in terms of threshold the effect is more pronounced for high values of $N_a$. Indeed, for both $e_1$ and $e_2$ the threshold when $N_a=5250$ is clearly better than for $N_a=5500$ if the correlation is not exploited, while $N_a=5500$ mostly outperforms $N_a=5250$ when the correlation is exploited.

Fig. \ref{fig:NP_sA8,8} also shows the excellent match between simulation results and DDE predictions in the error floor region when correlation is exchanged between the X and Z sections of the graph. As expected, due to the finite block length, the
convergence thresholds obtained through simulation are worse than
the DDE predictions. However, DDE predictions convey the same trend
as simulation results (better convergence thresholds as $N_a$ initially increases). The match between DDE and simulation results is also excellent when correlation is not exchanged between the X and Z sections of the graph, but the DDE curves for this case are not shown to avoid overcrowding the figure.

\begin{figure}[!t]
\centering
\hspace{-1cm}
\begin{minipage}{\textwidth}
\centering
\subfloat{\label{fig:fig1}\includegraphics[width=0.46\textwidth]{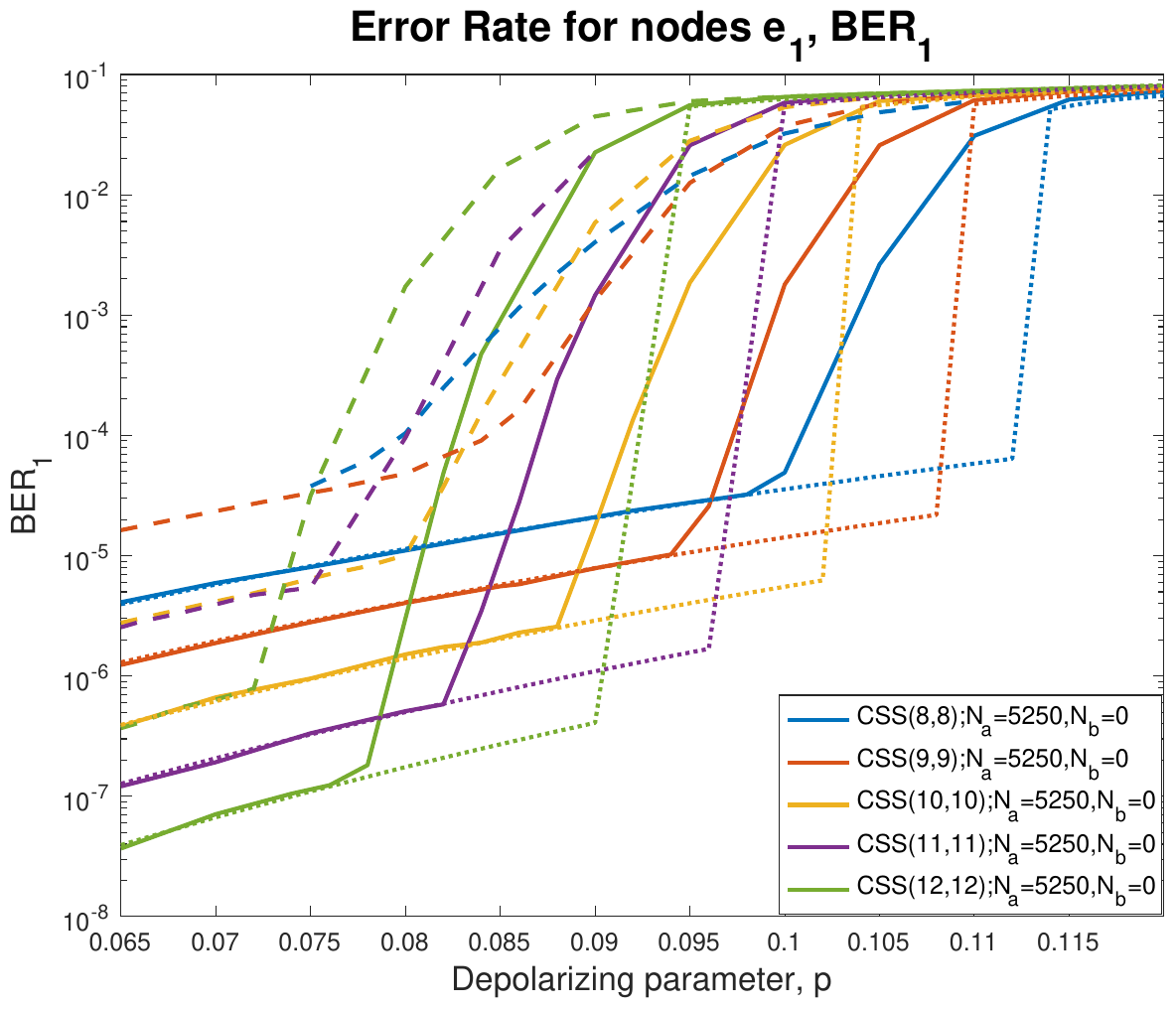}}
\hfill
\subfloat{\label{fig:fig2}\includegraphics[width=0.46\textwidth]{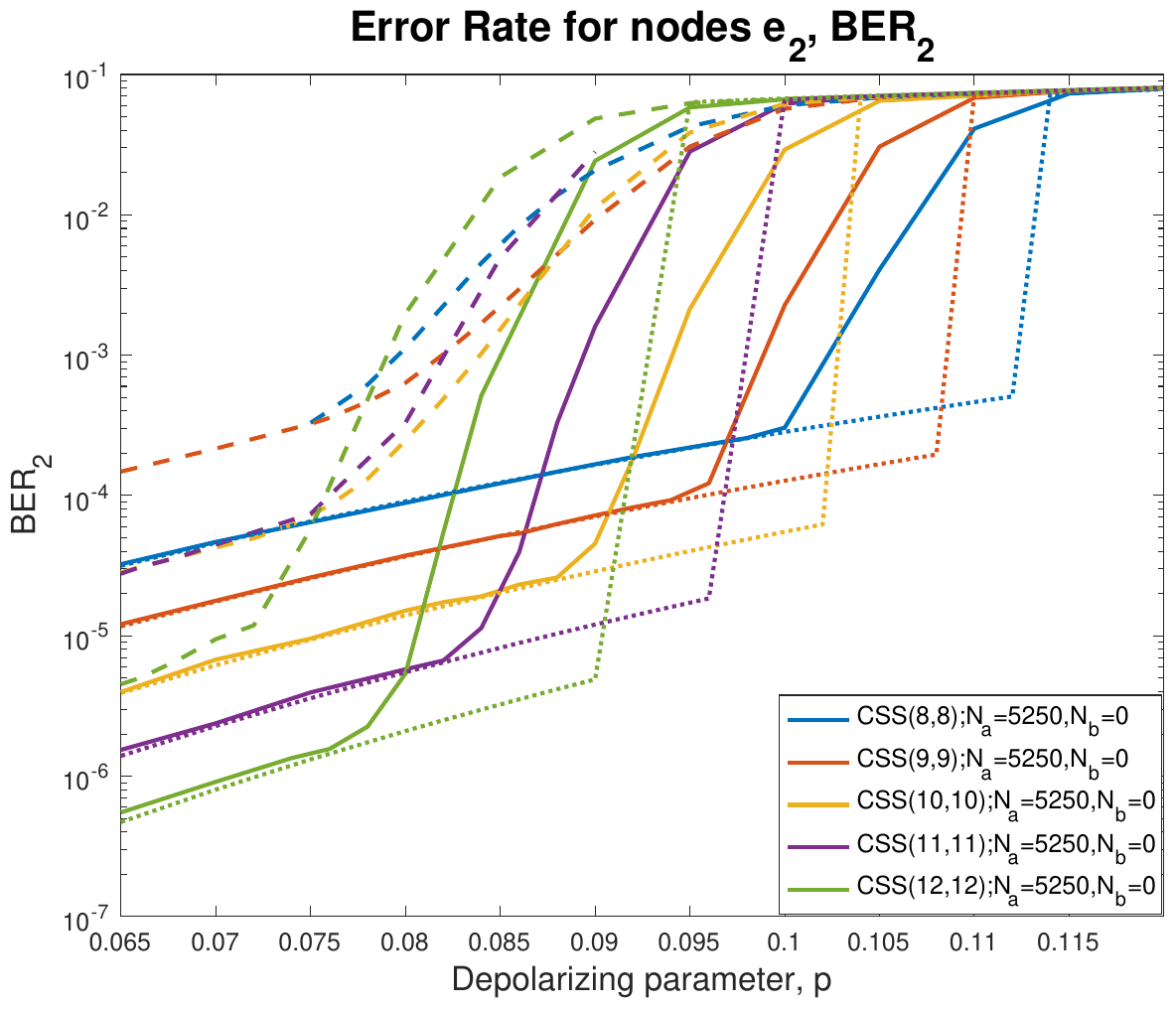}}
\end{minipage}
\caption{DDE predictions and simulated residual error rate in both $e_1$ and $e_2$ for a family of quantum codes designed using the original structure proposed in Fig. \ref{figure1} with an LDGM code of degrees varying from $(8,8)$ to $(12,12)$ in the bottom layer, $N_a=5250$ and $N_b=0$. A depolarizing channel is considered. The rightmost curves correspond to the case where the correlation between the $e^x$ and $e^z$ errors is exploited in the decoder via information exchange between the X and Z sections of the graph (continuous lines for simulation results and dotted lines for DDE predictions). The leftmost curves (dashed lines) correspond to the simulation results when this exchange does not occur.}
\vspace{-0.8cm}
\label{fig:NP_sA_BEST}
\end{figure}

Fig. \ref{fig:NP_sA_BEST} explores the effect of varying the degree of the LDGM code used in the bottom layer. $N_b$ is again fixed to 0, and $N_a=5250$, which resulted in the best performance in Fig. \ref{fig:NP_sA8,8}. As in Fig. \ref{fig:NP_sA8,8}, the rightmost curves correspond to the case where the correlation between the $e^x$ and $e^z$ errors is exploited in the decoder via information exchange between the X and Z sections of the graph (continuous lines for simulation results and dotted lines for DDE predictions), while the leftmost curves correspond to the case in which this exchange does not occur (dashed lines). Similar to Fig. \ref{fig:NP_sA8,8}, exploiting the correlation between the $e^x$ and $e^z$ errors (rightmost curves in the figure) leads to improvements in performance, both in terms of convergence thresholds and error floors, but the threshold improvement is more pronounced when the degree of the LDGM code is lower. As expected, in all cases (with and without correlation exploitation) the error floor decreases when the degree increases, while the convergence threshold experiences significant degradation. Notice also the excellent match between simulation results and DDE predictions in the error floor region when correlation is exchanged between the X and Z sections of the graph, and how DDE predictions convey the same trend
as simulation results. The match between DDE and simulation results is also excellent when correlation is not exchanged between the X and Z sections of the graph (results not shown).

Fig. \ref{fig:CSS_(8,8)} and Fig. \ref{fig:CSS_(12,12)} focus on designs using the original structure proposed in Fig. \ref{figure1} when the LDGM code in the bottom layer has degrees $(8,8)$ and $(12,12)$, respectively, and illustrate the effect of $N_b$ on the resulting performance. All curves are obtained exploiting the correlation between the $e^x$ and $e^z$ errors in the decoder via information exchange between the X and Z sections of the graph. The continuous lines correspond to the case where $N_a=5000$, while for the dashed lines $N_a=4000$. For both values of $N_a$ and in both figures ($(8,8)$ and $(12,12)$ codes), the convergence threshold improves when $N_b$ increases away from $0$. When $N_a=4000$, the best performance in both figures is obtained when $N_b=2500$, while for $N_a=5000$ the best results are achieved when $N_b=750$. In all cases, further increases of $N_b$ results in degradation of the convergence threshold. On the other hand, the error floor is independent of the value of $N_b$ (and $N_a$) and only depends on the degree of the LDGM codes. It is interesting to remark that for both the $(8,8)$ and the $(12,12)$ code, the scheme with $N_a=4000$ is much more sensitive to the value of $N_b$, as in this case the improvement in convergence threshold when $N_b$ increases away from $0$ is much more substantial than when $N_a=5000$, while for a fixed value of $N_a$ the $(8,8)$ code is more sensitive to changes in $N_b$. Notice that according to Fig. \ref{fig:CSS_(12,12)} it is possible to obtain codes with error floors around $10^{-6}$ and convergence thresholds at values of $p$ slightly above .08, while Fig. \ref{fig:CSS_(8,8)} indicates that codes with error floors around $10^{-4}$ and convergence thresholds at values of $p$ a little bit above .1 are possible.
\begin{figure}[!t]
\centering
\hspace{-1cm}
\begin{minipage}{\textwidth}
\centering
\subfloat{\label{fig:fig1}\includegraphics[width=0.46\textwidth]{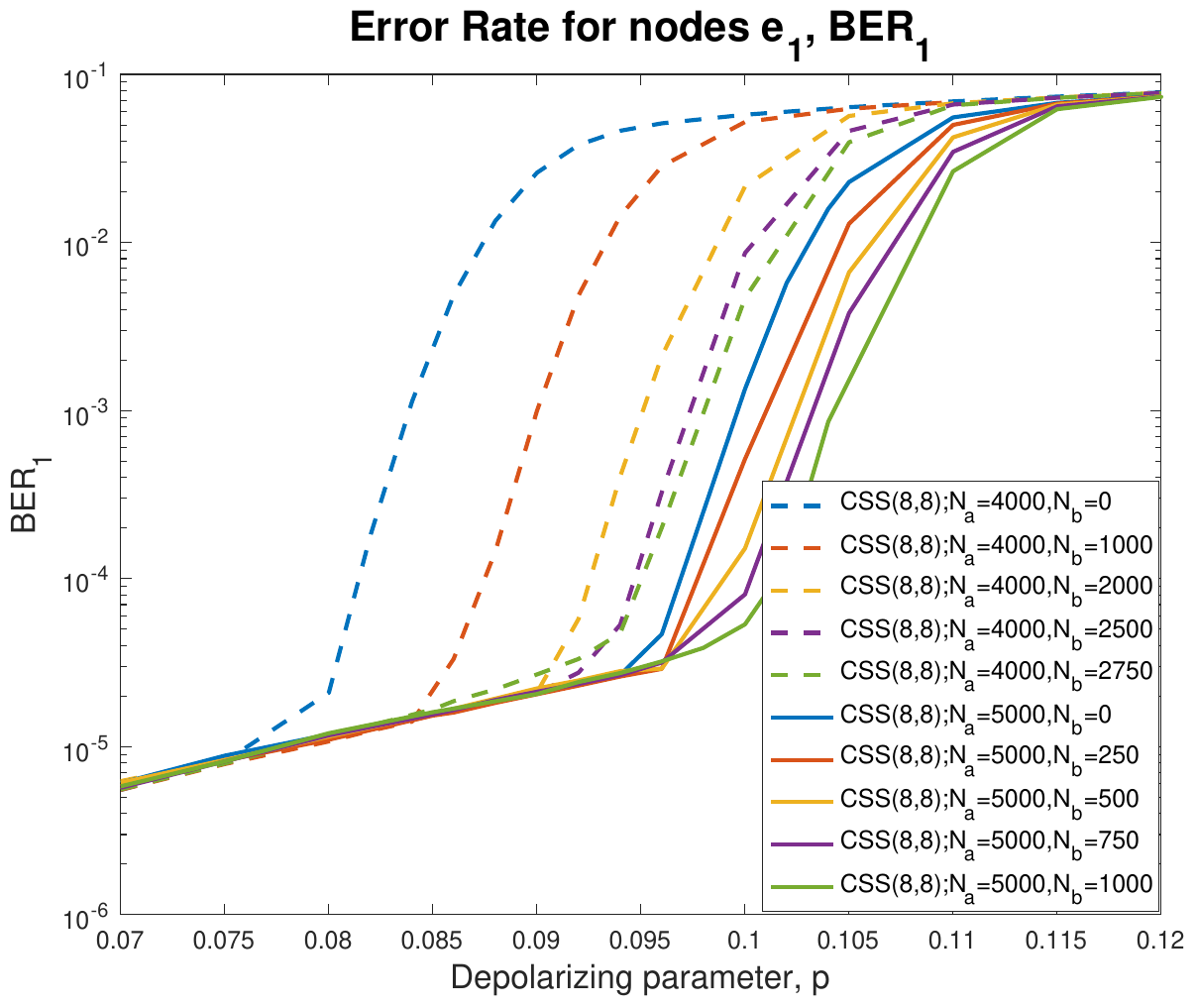}}
\hfill
\subfloat{\label{fig:fig2}\includegraphics[width=0.46\textwidth]{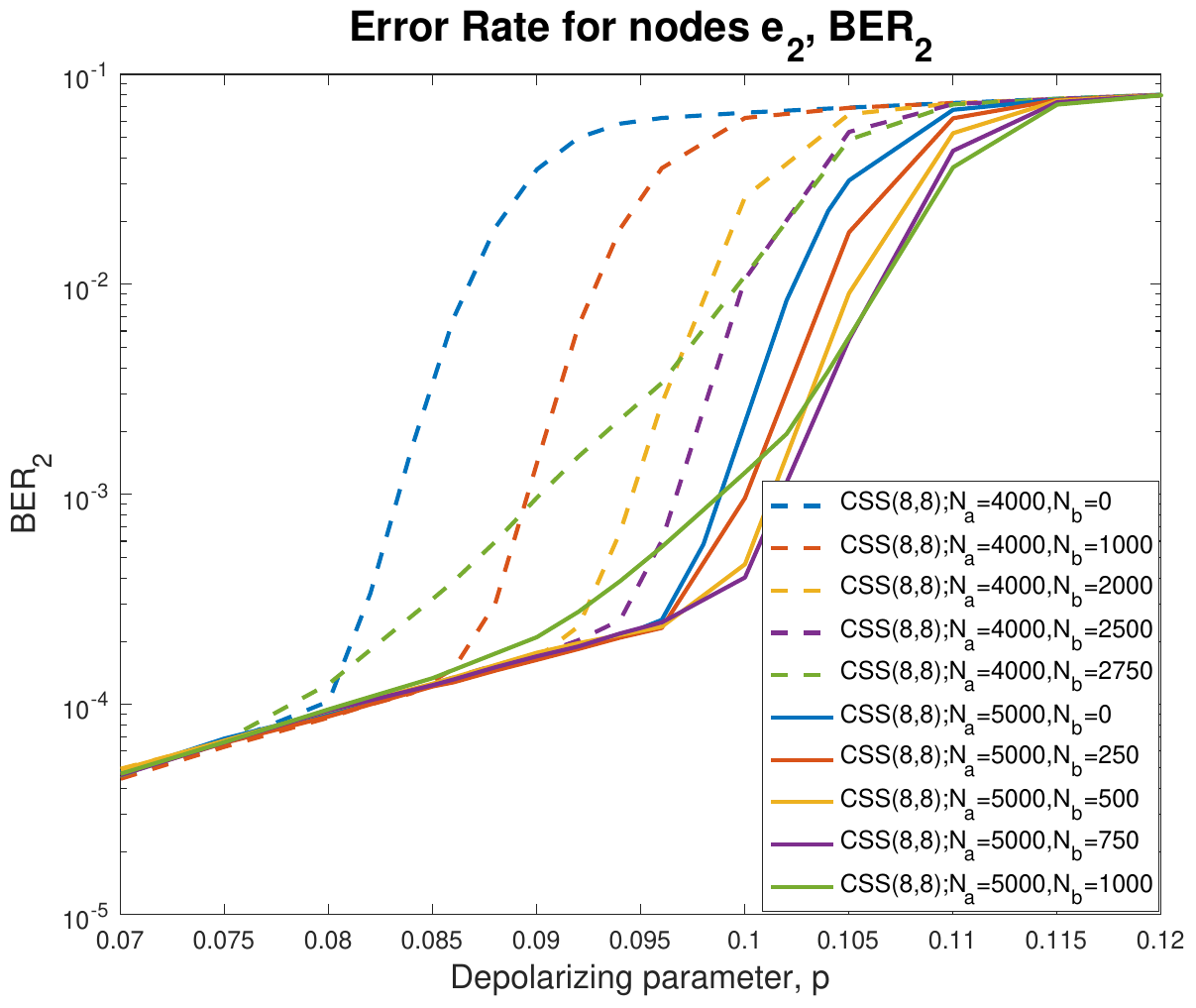}}
\end{minipage}
\caption{For CSS codes designed using the original structure proposed in Fig. \ref{figure1}, with an LDGM code of degrees $(8,8)$ in the bottom layer, simulated residual error rate in both $e_1$ and $e_2$ when $N_a=4000$ (dashed lines) and $N_a=5000$ (continuous lines) for different values of $N_b$. A depolarizing channel is considered, and in all cases the correlation between the $e^x$ and $e^z$ errors is exploited in the decoder via information exchange between the X and Z sections of the graph.}
\vspace{-0.8cm}
\label{fig:CSS_(8,8)}
\end{figure}

\begin{figure}[!t]
\centering
\hspace{-1cm}
\begin{minipage}{\textwidth}
\centering
\subfloat{\label{fig:fig1}\includegraphics[width=0.46\textwidth]{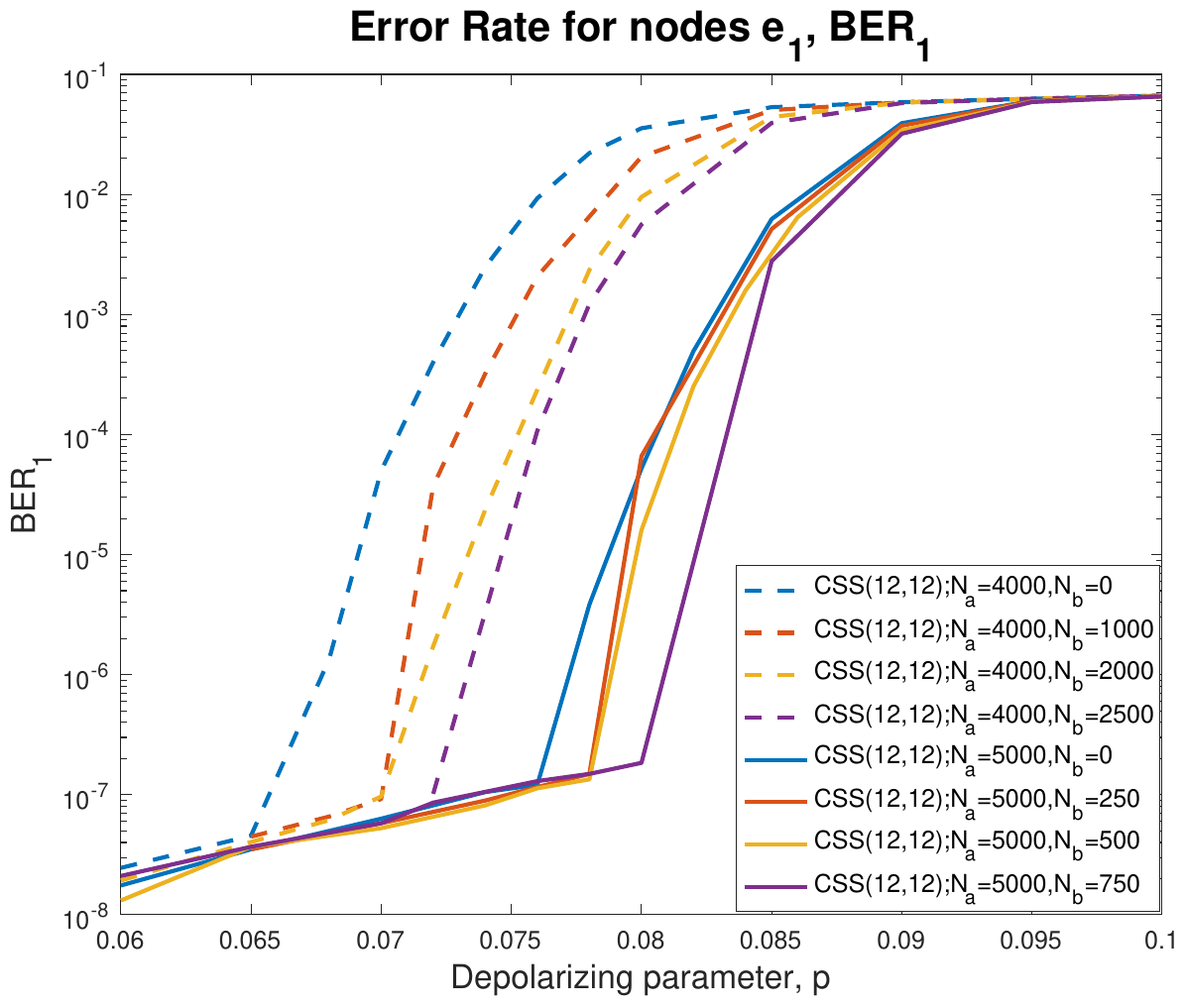}}
\hfill
\subfloat{\label{fig:fig2}\includegraphics[width=0.46\textwidth]{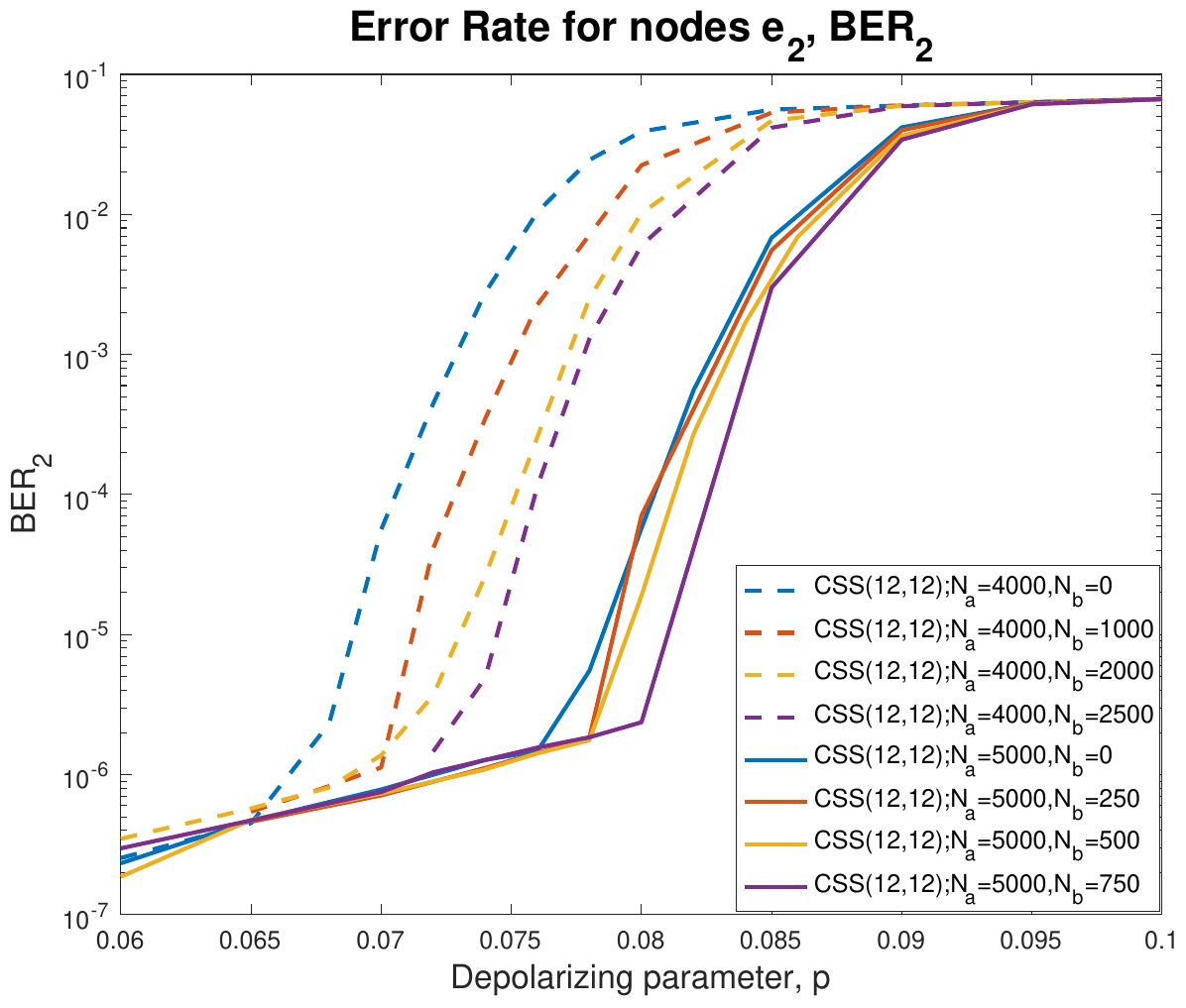}}
\end{minipage}
\caption{For CSS codes designed using the original structure proposed in Fig. \ref{figure1}, with an LDGM code of degrees $(12,12)$ in the bottom layer, simulated residual error rate in both $e_1$ and $e_2$ when $N_a=4000$ (dashed lines) and $N_a=5000$ (continuous lines) for different values of $N_b$. A depolarizing channel is considered, and in all cases the correlation between the $e^x$ and $e^z$ errors is exploited in the decoder via information exchange between the X and Z sections of the graph.}
\vspace{-0.8cm}
\label{fig:CSS_(12,12)}
\end{figure}

\begin{figure}[!t]
\centering
\hspace{-1cm}
\begin{minipage}{\textwidth}
\centering
\subfloat{\label{fig:fig1}\includegraphics[width=0.46\textwidth]{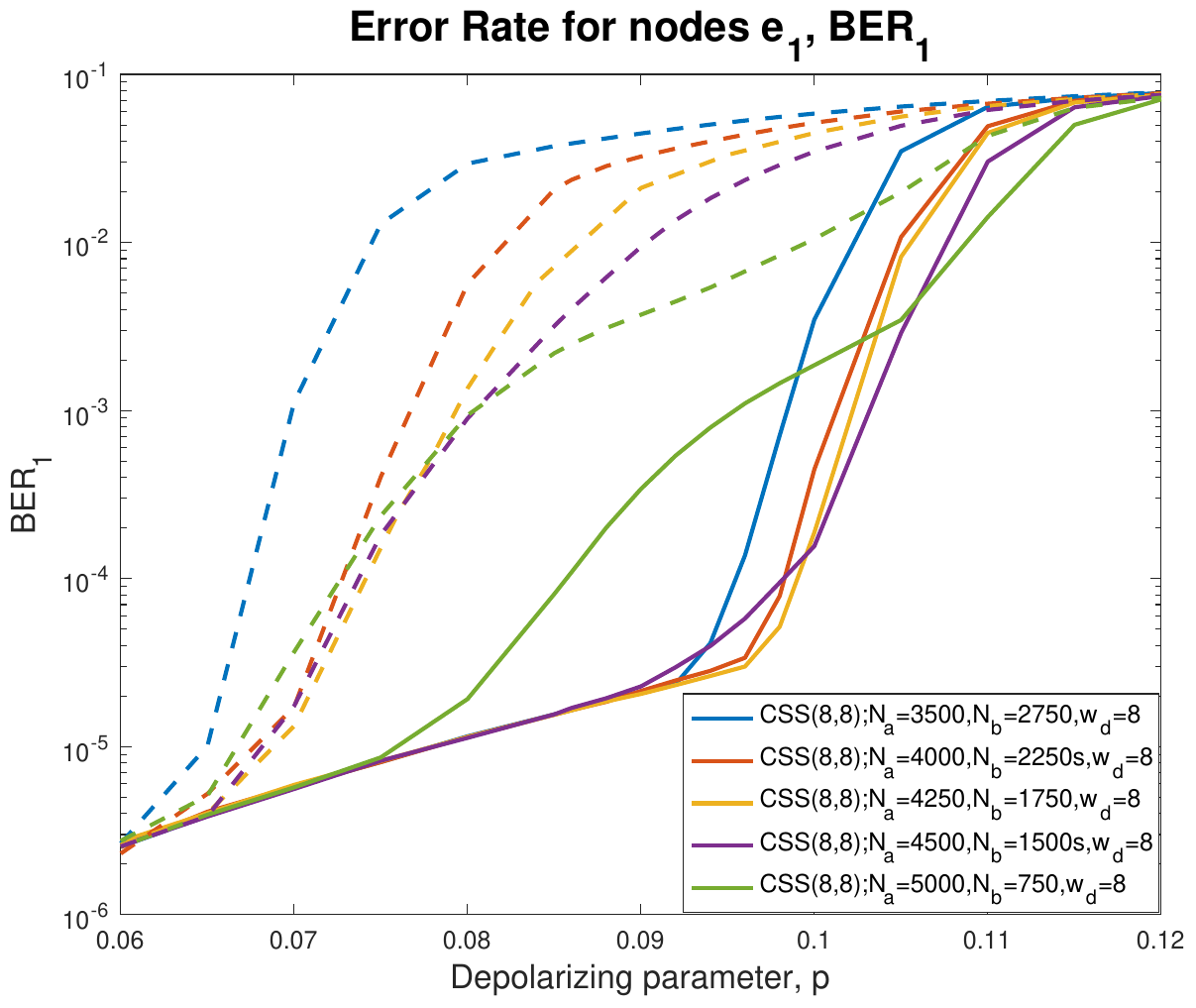}}
\hfill
\subfloat{\label{fig:fig2}\includegraphics[width=0.46\textwidth]{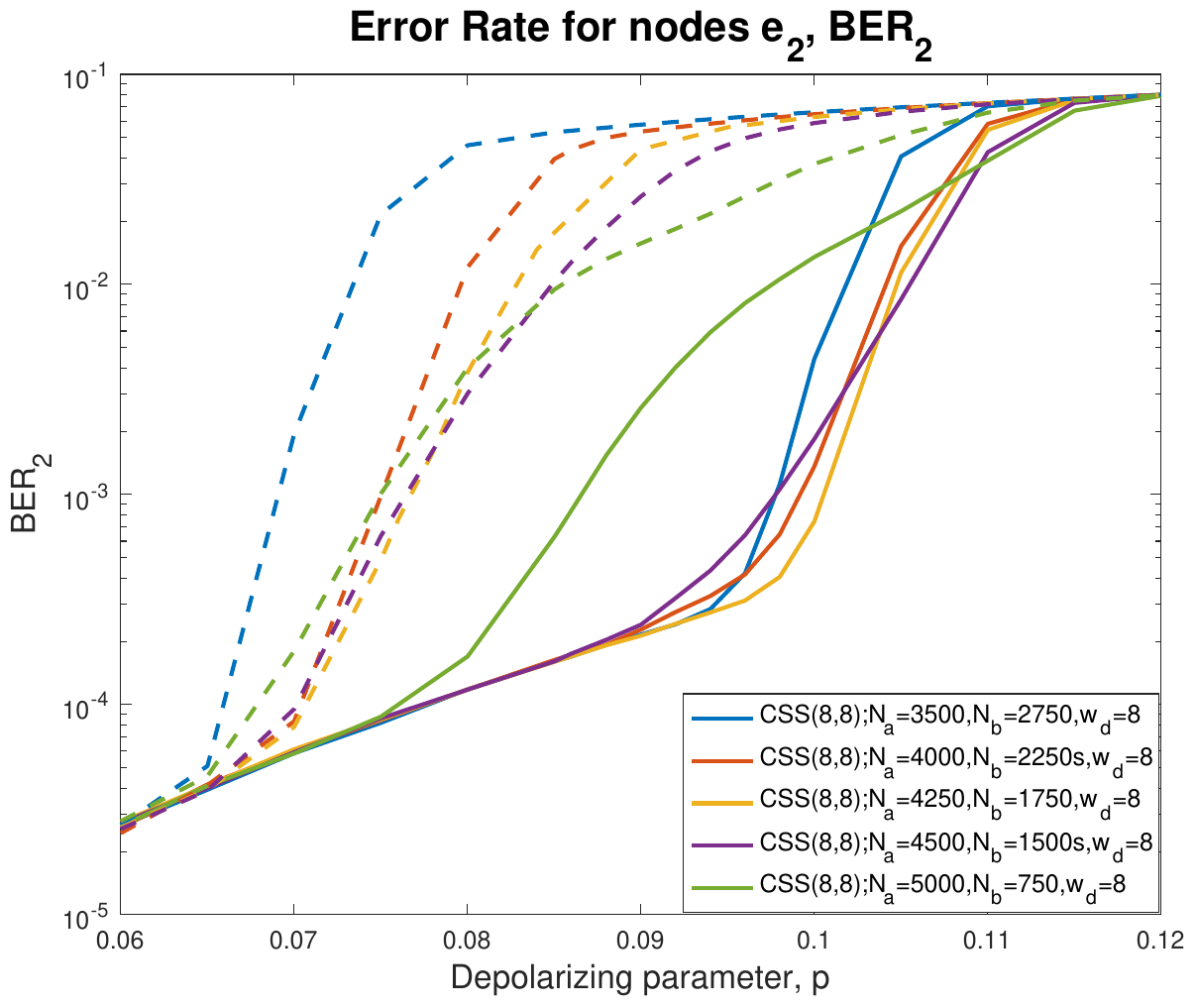}}
\end{minipage}
\caption{For CSS codes designed using the fault-tolerant structure proposed in Fig. \ref{figure3}, with an LDGM code of degrees $(8,8)$ in the bottom layer and $w_d=8$, simulated residual error rate in both $e_1$ and $e_2$ for different values of $N_a$ when i) $N_b$ is optimized for each value of $N_a$ (continuous lines; optimal value of $N_b$ provided in the label) and ii) $N_b=0$ (dashed lines). A depolarizing channel is considered, and in all cases the correlation between the $e^x$ and $e^z$ errors is exploited in the decoder via information exchange between the X and Z sections of the graph.}
\vspace{-0.8cm}
\label{fig:FT_(8,8)}
\end{figure}

\subsection{Fault-Tolerant Structure}
We now consider the design of codes well suited for fault-tolerant computation using the structure proposed in Section IV (see Fig. \ref{figure3}). Fig. \ref{fig:FT_(8,8)} considers a family of CSS codes with an LDGM code of degrees $(8,8)$ in the bottom layer when the degree of each node $d$ of degree different from 1 in the upper layer is $w_d=8$. All curves are obtained exploiting the correlation between the $e^x$ and $e^z$ errors in the decoder via information exchange between the X and Z sections of the graph. For a given value of $N_a$, the continuous (rightmost) curves indicate the performance for the optimal value of $N_b$, which is identified in the label. The dashed (leftmost) lines represent the performance for each value of $N_a$ when $N_b=0$. It is important to remark that, similar to the $(8,8)$ code designed in Fig. \ref{fig:CSS_(8,8)} using the original structure, optimizing $N_b$ leads to substantial improvements in the convergence threshold while the error floor is not affected. This improvement is particularly relevant for smaller values of $N_a$. Notice that these codes achieve error floors a little bit higher than $10^{-4}$ and convergence thresholds at values of $p$ around .1, very similar to those presented in  Fig. \ref{fig:CSS_(8,8)} using the original structure with an $(8,8)$ code in the bottom layer.

\begin{figure}[!t]
\centering
\hspace{-1cm}
\begin{minipage}{\textwidth}
\centering
\subfloat{\label{fig:fig1}\includegraphics[width=0.46\textwidth]{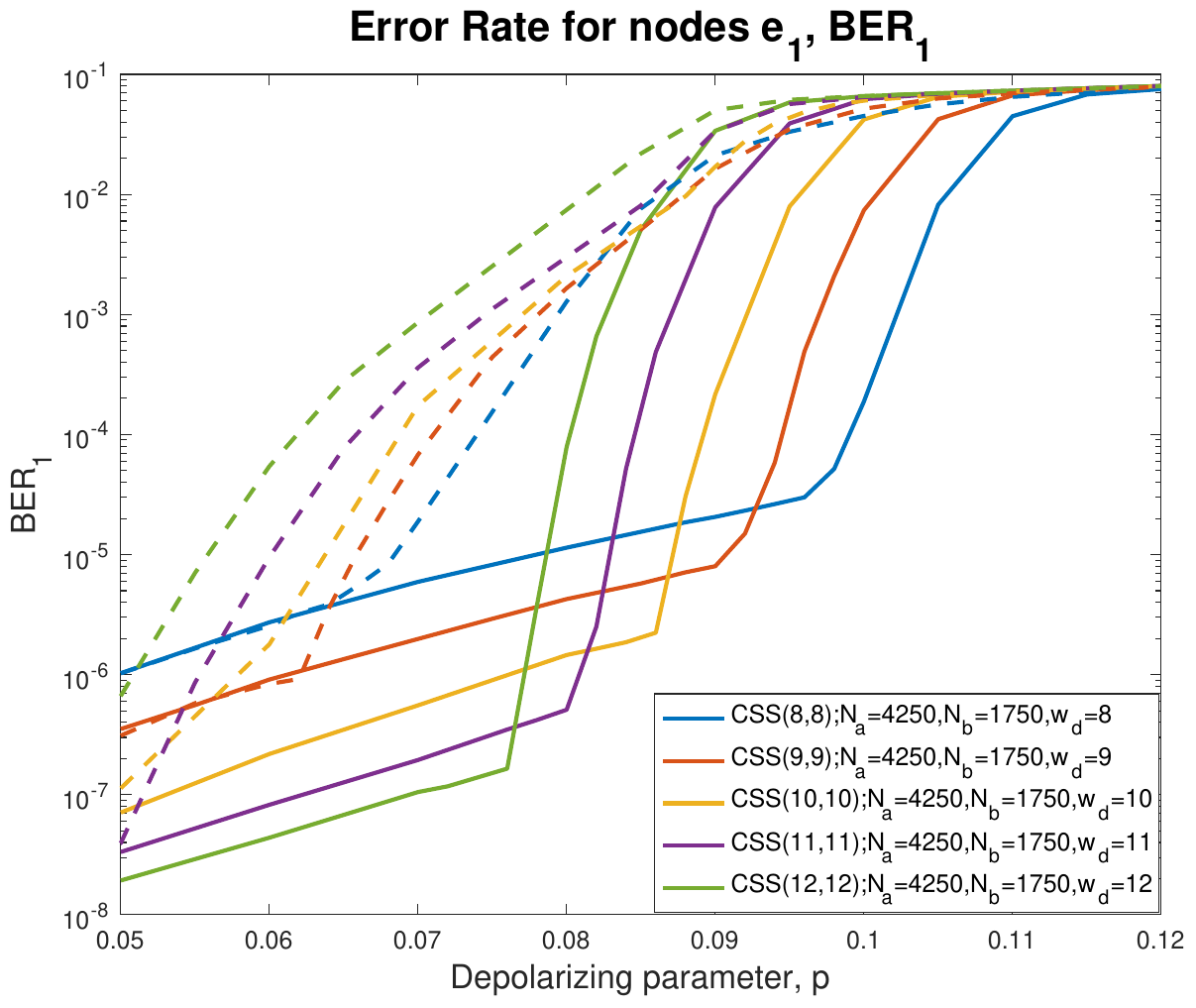}}
\hfill
\subfloat{\label{fig:fig2}\includegraphics[width=0.46\textwidth]{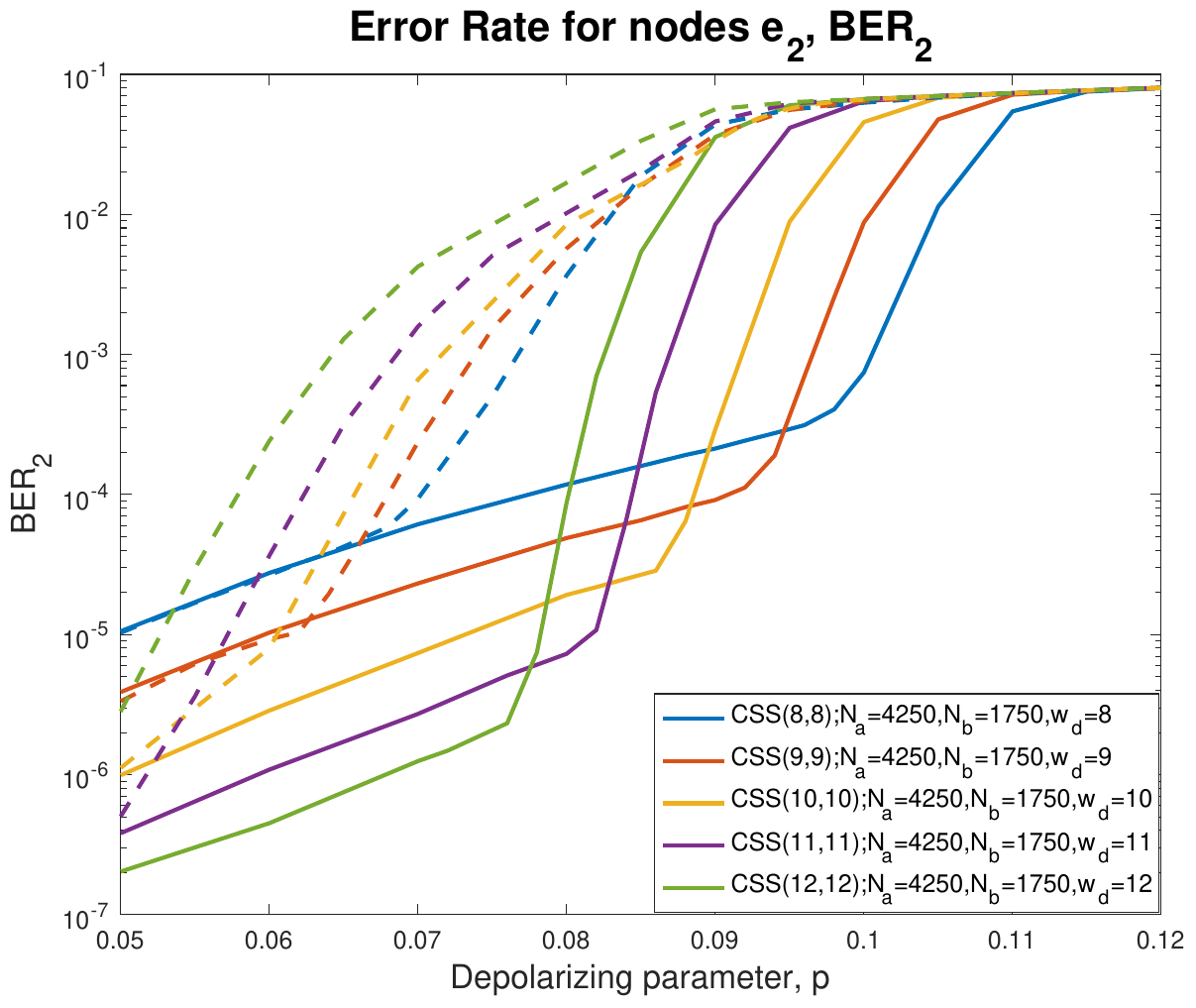}}
\end{minipage}
\caption{Simulated residual error rate in both $e_1$ and $e_2$ for a family of quantum codes designed using the fault-tolerant structure proposed in Fig. \ref{figure3}, with an LDGM code of degrees $(y,y),y \in \{8,9,10,11,12 \}$ in the bottom layer and $N_a=4250$, $w_d=y$ in the upper layer. Two cases are studied: i) $N_b$ is optimized for each value of $N_a$ (continuous lines; optimal value of $N_b$, always $1750$, provided in the label) and ii) $N_b=0$ (dashed lines). A depolarizing channel is considered, and in all cases the correlation between the $e^x$ and $e^z$ errors is exploited in the decoder via information exchange between the X and Z sections of the graph.}
\vspace{-0.8cm}
\label{fig:FT_BEST}
\end{figure}
Fig. \ref{fig:FT_BEST} explores the effect of varying the degree of the LDGM code used in the bottom layer, $(y,y),y \in \{8,9,10,11,12 \}$  when $N_a=4250$ and $w_d=y$ in the upper layer. All curves are obtained exploiting the correlation between the $e^x$ and $e^z$ errors in the decoder via information exchange between the X and Z sections of the graph. The continuous (rightmost) curves indicate the performance of each code for the optimal value of $N_b$, which is identified in the label (always 1750). The dashed (leftmost) lines represent the performance for each one of the codes when $N_b=0$. In all cases, optimizing $N_b$ leads to substantial improvements in the convergence threshold while the error floor is not affected. This improvement is more relevant for lower values of $w_d$. Notice the trade-off between the error floor, which decreases as the degree increases, and the convergence threshold, which degrades as the degree increases. This makes it possible to obtain codes with error floors around $10^{-6}$ and convergence thresholds at values of $p$ slightly below .08 for LDGM codes with degrees $(12,12)$, but also error floors a little bit higher than $10^{-4}$ and convergence thresholds at values of $p$ around .1 when the degree is $(8,8)$.

\section{Conclusion}
We have constructed a new family of Calderbank-Shor-Steane (CSS) codes using 
the generator and parity-check matrices of Low-Density Generator Matrix (LDGM) codes concatenated with ``compression" matrices that perform  row operations
to achieve the desired quantum rate. For the depolarizing channel, decoding is 
performed in an iterative manner, by applying message passing and properly exploiting the correlations existing in the associated decoding graph. 

We have also developed discrete Density Evolution (DDE) to predict the performance of the proposed codes in the depolarizing channel. Remarkably, DDE predictions match very well with simulation results, and both show that exploiting the correlations existing in the decoding graph leads to significant improvements in the error-correction capability, both in terms of convergence threshold and error floors. Although density evolution is well known in classical coding, to our knowledge this is the first time it has been applied in quantum codes over the depolarizing channel. The challenge is that the classical graph utilized in density evolution to study the performance of quantum codes does not have a direct physical meaning in the quantum domain. However, in terms of predictive performance this graph is equivalent to the one utilized for the iterative decoding of the quantum code, and therefore the resulting predictions apply to the proposed quantum codes.

The proposed construction offers high flexibility and easiness in the design, producing quantum codes with excellent error correction capabilities. By properly constructing the parity-check matrix of the code, we are able to control and bound the weight of the encoded/logical $\bar{X}$ and $\bar{Z}$ operators to a small value. This allows the design of codes particularly well suited for fault-tolerant quantum computation, which, at the same time, possess excellent error correction capabilities, with a performance only slightly worse than that of the best codes obtained when the only focus is on minimizing the residual errors.
\section{Contribution}
HL and JGF conceived and developed the original structure (Section II). YL developed the doping scheme (Section III) and conceived and designed the fault-tolerant codes (Section IV). KL, YL, and JGF developed the discrete density evolution framework (Section VI); the latter two also adapted the density evolution and decoding algorithms (Section V) for the depolarizing channel. Finally, YL ran the simulations (Section VII). All authors contributed to writing the manuscript, and JGF supervised the project.


\begin{thebibliography}{99}


\bibitem{shor95} P. Shor, ``Scheme for Reducing Decoherence in Quantum
Computer Memory,'' {\it Physical Review A}, vol. 52 (R), no. 4, pp.
2493-2496, October 1995.



\bibitem{NCbook} M. A. Nielsen and I. L. Chuang, ``Quantum
Computation and Quantum Information,'' {\it Cambridge University
Press,} 2000.



\bibitem{cs96} R. Calderbank and P. W. Shor, ``Good Quantum
Error-Correcting Codes Exist,'' {\it Physical Review A}, vol. 54,
no. 2, pp. 1098-1105, August 1996.



\bibitem{steane96} A. Steane, ``Multiple Particle Interference and
Quantum Error Correction,'' {\it Proc. of the Royal Society A}, vol.
452, no. 1954, pp. 2551-2577, November 1996.


\bibitem{gottesmanThesis}
D.~Gottesman,
\newblock \emph{Stabilizer Codes and Quantum Error Correction},
\newblock Ph.D. thesis, California Institute of Technology, 1997.
\newblock Available at \url{https://arxiv.org/abs/quant-ph/9705052}.

\bibitem{CRSS}
A.~R. Calderbank, E.~M. Rains, P.~W. Shor, and N.~J.~A. Sloane,
\newblock Quantum error correction and orthogonal geometry,
\newblock \emph{Phys. Rev. Lett.} \textbf{78}(3), 405–408 (1997).

\bibitem{crss98} A. R. Calderbank, E. M. Rains, P. W. Shor, and N.
J. A. Sloane, ``Quantum Error Correction via Codes over GF(4),''
{\it IEEE Trans. on Information Theory}, vol. 44, no. 4, pp.
1369-1387, July 1998.

M.~M. Wilde,
\newblock \emph{Quantum Information Theory}, 2nd ed.
\newblock Cambridge University Press, 2017.

\bibitem
{berrou}
C. Berrou, A. Glavieux, and P. Thitimajshima, ``Near Shannon Limit
Error-Correcting Coding and Decoding: Turbo-Codes,'' {\it Proc. ICC'93},
May 1993.

\bibitem{gallager} R. G. Gallager, ``Low-Density Parity-Check
Codes,'' {\it IEEE Trans. on Information Theory}, vol. 8, no. 1, pp.
21-28, January 1962.



\bibitem{mackay1999} D. J. C. MacKay, ``Good Error-Correcting Codes
Based on Very Sparse Matrices,'' {\it IEEE Trans. on Information
Theory}, vol. 45, no. 2, pp. 399-431, March 1999.

\bibitem{postol}
M. S. Postol, ``A Proposed Quantum Low Density Parity Check Code,''
{\it arXiv:quant-ph/0108131v1}, August 2001.

\bibitem{mackay} D. J. C. MacKay, G. Mitchison, and P. L. McFadden,
``Sparse-Graph Codes for Quantum Error-Correction,'' {\it IEEE
Trans. on Information Theory}, vol. 50, no. 10, pp. 2315-2330,
October 2004.

\bibitem{camara}
T. Camara, H. Ollivier and J.-P. Tillich, ``Constructions and
Performance of Classes of Quantum LDPC Codes,'' {\it
arXiv:quant-ph/0502086v2}, April 2005.

\bibitem{spawc05} H. Lou and J. Garcia-Frias, ``Quantum
Error-Correction Using Codes with Low-Density Generator Matrix,''
{\it Proc. SPAWC'05}, June 2005.

\bibitem{turbo06} H. Lou and J. Garcia-Frias, ``On the Application
of Error-Correcting Codes with Low-Density Generator Matrix over
Different Quantum Channels,'' {\it Proc. International Symposium on
Turbo Codes}, April 2006.

\bibitem{ciss08} J. Garcia-Frias and K. Liu, ``Design of
Near-Optimum Quantum Error-Correcting Codes Based on Generator and
Parity-Check Matrices of LDGM Codes,'' {\it Proc. CISS'08}, March
2008.

\bibitem{gz2003} J. Garcia-Frias and W. Zhong, ``Approaching Near
Shannon Performance by Iterative Decoding of Linear Codes with
Low-Density Generator Matrix,'' {\it IEEE Communications Letters,}
vol. 7, no. 6, pp. 266-268, June 2003.

\bibitem{asil} J. Garcia-Frias, W. Zhong, and Y. Zhao, ``Iterative
Decoding Schemes for Source and Channel Coding of Correlated
Sources,'' {\it Proc. Asilomar'02}, November 2002.

\bibitem{Wei} W. Zhong and J. Garcia-Frias, ``LDGM Codes for Channel
Coding and Joint Source-Channel Coding of Correlated Sources,'' {\it
EURASIP Journal on Applied Signal Processing}, vol. 2005, no. 6, pp.
942-953, May 2005.

\bibitem{Rengaswamy}
N.~Rengaswamy, R.~Calderbank, S.~Kadhe, and H.~D.~Pfister,
``Synthesis of Logical Clifford Operators via Symplectic Geometry,''
in \emph{Proc. IEEE Int. Symp. Inf. Theory (ISIT)}, 2018, pp.~791--795.
doi: 10.1109/ISIT.2018.8437652.

\bibitem{Rengaswamy2}
N.~Rengaswamy, R.~Calderbank, S.~Kadhe, and H.~D.~Pfister,
``Synthesis of Logical Clifford Operators via Symplectic Geometry,'' \emph{http://arxiv.org/abs/1803.06987}, 2018.

\bibitem{Panteleev_SIGACT}
P.~Pantellev and G~Kalachev,
``Asymptotically Good Quantum and Locally Testable Classical LDPC Codes,'' in \emph{Proc. ACM SIGACT Symposium on Theory of Computing}, 2022, pp.~375--388. doi: 10.1145/3519935.3520017.

\bibitem{Panteleev_IT}
P.~Pantellev and G~Kalachev,
``Quantum LDPC Codes With Almost Linear Minimum Distance,'' {\it IEEE
Trans. on Information Theory}, vol.68, no. 1, pp. 213--229,
January 2022. doi: 10.1109/TIT.2021.3119384. 


\bibitem{golowich} L.~Golowich and V.~Guruswami, ``Quantum LDPC Codes of Almost Linear Distance via Homological Products,'' {\it arXiv:quant-ph/2411.03646 }, November 2024.

\bibitem{liu_isit10} K. Liu, J. Garcia-Frias, ``Error Floor Analysis in LDGM
Codes'', {\it Proc. ISIT'10}, June 2010.


\bibitem{chai05} H. Chai, W. Zhong, and J. Garcia-Frias, ``Parallel
Concatenation of LDGM Codes to Approach Capacity Limits,'' {\it
Proc. CISS'05}, March 2005.

\bibitem{isit05} W. Zhong, H. Chai, and J. Garcia-Frias, ``Approaching
the Shannon Limit through Parallel Concatenation of Regular LDGM
Codes,'' {\it Proc. ISIT'05}, September 2005.



\bibitem{dde}
S.-Y. Chung, G. D. Forney, T. J. Richardson, and R. Urbanke,  ``On
the Design of Low-Density Parity-Check Codes within 0.0045 dB of the
Shannon Limit,'' {\it IEEE Communications Letters}, vol. 5, no. 2,
pp. 58-60, February 2001.

\bibitem{pearl1988} J. Pearl, ``Probabilistic Reasoning in
Intelligent Systems: Networks of Plausible Inference,'' {\it Morgan
Kaufmann}, 1988.

\bibitem{kschischang2001} F. R. Kschischang, B. J. Frey, and H.-A.
Loeliger, ``Factor Graphs and the Sum-Product Algorithm,'' {\it IEEE
Trans. on Information Theory}, vol. 47, no. 2, pp. 498-519, February
2001.

\bibitem{almeida}
A. C. A. de Almeida and R. Palazzo Jr., ``A Concatenated [(4,1,3)]
Quantum Convolutional Code,'' {\it Proc. ITW'04}, October 2004.

\bibitem{forney}
G. D. Forney, Jr., M. Grassl and S.Guha, ``Convolutional and
Tail-Biting Quantum Error-Correcting Codes,'' {\it
arXiv:quant-ph/0511016v2}, November 2006.

\bibitem{grassl1}
M. Grassl and T. Beth, ``Quantum BCH Codes,'' {\it
arXiv:quant-ph/9910060v1}, October 1999.

\bibitem{grassl2}
M. Grassl and T. Beth, ``Quantum Reed-Solomon Codes,'' {\it
arXiv:quant-ph/9910059v1}, October 1999.

\bibitem{grassl3}
M. Grassl and M. Roetteler, ``Constructions of Quantum Convolutional
Codes,'' {\it Proc. ISIT'07}, June 2007.

\bibitem{hagiwara}
M. Hagiwara, and H. Imai, ``Quantum Quasi-Cyclic LDPC Codes,'' {\it
Proc.  ISIT'07}, June 2007.

\bibitem{ollivier}
H. Ollivier and J.-P. Tillich, ``Description of a Quantum
Convolutional Code,'' {\it Phys. Rev. Lett.}, vol. 91, no. 17, pp.
177902-1-4, October 2003.

\bibitem{doping} S. ten Brink, ``Code Doping for Triggering
Iterative Decoding Convergence,'' {\it Proc. ISIT'01}, June 2001.

\bibitem{ciss09} K. Liu and J. Garcia-Frias, ``Asymptotic Analysis of LDGM-Based Quantum Codes,'' {\it Proc. CISS'09}, March
2009.

\bibitem
{dcc01}
J. Garcia-Frias, ``Joint Source-Channel Decoding of Correlated Sources 
over Noisy Channels,'' {\it Proc. DCC'01}, March 2001.

\bibitem
{tcom_noise}
J. Garcia-Frias and Y. Zhao, ``Near Shannon/Slepian-Wolf Performance for Unknown Correlated Sources over AWGN Channels,'' {\it IEEE Trans. on
Communications}, vol. 53, no. 4, pp. 555-559, April 2005.

\bibitem
{cl}
J. Garcia-Frias and Y. Zhao, ``Compression of Correlated Binary Sources
Using Turbo Codes,'' {\it IEEE Communications Letters}, vol. 5, no. 10, 
pp. 417-419, October 2001.

\bibitem
{spm07} J. Garcia-Frias, Y. Zhao, and W. Zhong, ``Turbo-Like Codes for Transmission of Correlated Sources over Noisy Channels,'' {\it IEEE Signal Processing Magazine}, vol. 24, no. 5, pp. 58-66, September 2007.



\bibitem{Hagenauer} J. Hagenauer, E. Offer and L. Papke ``Iterative decoding
of binary block and convolutional codes,'' {\it IEEE Trans. Inform.
Theory}, vol. 42, pp. 429-445, March 1996.





\end{thebibliography}
\end{document}